\documentclass[12pt]{article}        
\usepackage{axodraw}
\usepackage{epsfig}
\usepackage{cite}

\renewcommand{\thefootnote}{\fnsymbol{footnote}}

\begin{document}                     


\begin{titlepage}
\date{date} 
\begin{flushright}
{\bf  Budker INP 2002-2\\
January 2002}
\end{flushright}
\vspace{1mm}
\begin{center}
\vspace{1cm}

{\LARGE \bf Novosibirsk hadronic currents} \\
\vspace{3mm}
{\LARGE \bf for  $\tau \to 4\pi$  channels of  $\tau$ decay library}\\
\vspace{3mm}
{\LARGE \bf TAUOLA~\footnote[2]{~Work supported in part by the Polish 
State Committee grants KBN 
5P03B09320, 
5P03B10721 
and NATO Grant PST.CLG.977751, European
 Commission 5-th Framework contract HPRN-CT-2000-00149.}}\\
\vspace{0.1cm}

\end{center}

\vspace{0.3cm}
\begin{center}
{\large \bf  A. E. Bondar$^{a}$, S. I. Eidelman$^{a}$, A. I. Milstein$^{a}$,}\\

\vspace{2mm}
{\large \bf  T. Pierzcha\l a$^{b}$, N. I. Root$^{a}$, Z. W\c{a}s$^{c,d}$ and  M. Worek$^{b}$ }
\end{center}

\vspace{5mm}
\begin{center}

$^{a}$ {\em Budker Institute of Nuclear Physics\\ Academician Lavrentyev 11,
630090 Novosibirsk-90, Russia}\\
$^{b}${\em Institute of  Physics, University of Silesia\\ Uniwersytecka 4,
 40-007, Katowice, Poland}\\
$^c${\em Institute of Nuclear Physics\\
 Kawiory 26a, 30-055 Cracow, Poland}\\
$^d${\em CERN, Theory Division, 1211 Geneva 23, Switzerland}

\end{center}

\vspace{1cm}
\begin{abstract}
 The new parameterization of form factors developed for 4$\pi$ channels of
the $\tau$ lepton decay and based on Novosibirsk data 
on $e^+e^- \to 4\pi$ has been coded in 
a form suitable for the {\tt TAUOLA} Monte Carlo package.  Comparison with 
results from {\tt TAUOLA} using another parameterization, {\it i.e.}  the
CLEO version of 1998 is also included.
\end{abstract} 

\vspace{ 0.5 cm}
\begin{center}
{\it To be submitted to Comput. Phys. Commun.} 
\end{center}
\vfill

\end{titlepage}
\renewcommand{\thefootnote}{\arabic{footnote}}
\section{Introduction}
\vskip 0.3 cm 

Hadronic decays of the $\tau$ lepton offer a unique laboratory for studying
hadronic interactions at low energies (below the $\tau$ lepton mass
$M$). Proper 
modeling of such processes will facilitate comparison of experiment and
theoretical models and may thus  offer a crucial hint toward better 
understanding 
of low energy phenomenology of strong interactions. 
In this work we consider the $\tau$ decay into four pions and neutrino.
Several models of this decay are at present in use 
\cite{Pions1,Pions2,CLEO2000,CLEO,Binp1,Binp2}. While 
Refs.~\cite{Pions1,Pions2} make an attempt to construct 
the Lagrangian of the decay from general theoretical principles,
other models are purely phenomenological using
either the experimental information from $\tau$ decays like
in Refs.~\cite{CLEO2000,CLEO} or from $e^+e^-$ annihilation 
like in Refs.~\cite{Binp1,Binp2}.  
In the following let us concentrate on the latter approach.

Production of four pions is one of the dominant processes of $e^{+}e^{-}$
annihilation into hadrons in the energy range from $1.05$ to $2.5$ GeV. 
The hypothesis of the conserved vector current (CVC) relates to each other 
the cross section of this process and 
$\tau\rightarrow4\pi\nu$ decay \cite{CVC}. Therefore, all realistic models 
which describe the first process should also properly describe the other one. 
For a recent review of theoretical predictions for 
various decay modes of the $\tau$ based on CVC see {\it e.g.}
\cite{CVC-test}.  

For more than five years, two new 
detectors CMD-2 \cite{CMD-2} and SND \cite{SND}
have been studying low energy $e^{+}e^{-}$ annihilation at the $e^{+}e^{-}$ 
collider VEPP-2M at Novosibirsk. Their results can be used to provide
a new parameterization of the form factors used in the Monte Carlo generators 
of the $\tau$ lepton.
In the present paper we define  the form factors which can be used as 
one of the possibilities in
{\tt TAUOLA} \cite{tauola:1990,tauola:1992,tauola:1993}, the $\tau$ 
decay library. The {\tt TAUOLA} package is organized in such a way that 
the phase space generation and calculation of the electroweak part of 
the matrix element are separated from the part of the code calculating 
the hadronic current provided by the particular model.
This is convenient not only for constructing the Monte Carlo program, 
but  also facilitates comparisons between different models.
The technical description of the solution which allows an easy replacement 
of different hadronic currents is described in 
\cite{golonka:2000,pierzchala:2001}.

Our paper is organized as follows: in the next section we recall basic 
principles of how $e^+e^-$ data can be
used in defining hadronic currents for the $\tau$ lepton decay. 
In Section 3 we describe  those parts of the {\tt TAUOLA} Monte Carlo
algorithm for four pion generation and general formalism for the  
semileptonic/semihadronic decays which are needed for the definition 
of hadronic currents. 
In Section 4 we describe the new 
current for the $4\pi$ decay of the $\tau$ while in Section 5 
details of the functions and constants used in this work are presented.
Section 6 is devoted to technical tests of our code. 
Section 7 shows the comparison of 
numerical results from running the new version of {\tt TAUOLA}
with the relatively old one based on the CLEO  parameterization
 \cite{CLEO_OLD,test_cleo1,test_cleo2,test_cleo3} 
which is still widely in use  despite the recent 
significant progress achieved by CLEO \cite{CLEO,CLEO2000}. 
Chapter 8 briefly summarizes the main results of the paper.

In Appendices  we describe some technical aspects of our
model. In particular, we tabulate
the functions $G(Q^{2})$ used for the definition of the hadronic current.

\section{ Relation between $\tau$ decays and $e^{+}e^{-}$ annihilation
cross sections }
\vskip 0.3 cm

Via the hypothesis of conserved vector current one can relate 
the charged vector current coupling to the $4\pi$ system  
to the electromagnetic (neutral vector) current measured by 
$\sigma(e^{+}e^{-}\rightarrow\gamma\rightarrow4\pi)$.
There are two possible final states in $e^{+}e^{-}$ annihilation
\footnote{ ~The $e^{+}e^{-}\rightarrow\pi^{0}\pi^{0}\pi^{0}\pi^{0}$ channel is 
forbidden by isospin and charge 
conjugation invariance.}
\[
e^{+}e^{-}\rightarrow\gamma^{*}\rightarrow\tilde{\rho}^{0}\rightarrow
\pi^{-}\pi^{-}\pi^{+}\pi^{+},
\]
\begin{equation}
e^{+}e^{-}\rightarrow\gamma^{*}\rightarrow\tilde{\rho}^{0}\rightarrow\pi^{+}
\pi^{-}\pi^{0}\pi^{0},
\label{ee}
\end{equation} 
They are accessible by a different $I_{3}$ 
component of the same $I=1$ weak
 current describing $\tau$ decay:
\[
\tau^{+}\rightarrow W^{+}\bar{\nu}_{\tau} \rightarrow \tilde{\rho}^{+}
\bar{\nu}_{\tau}\rightarrow
\bar{\nu}_{\tau}\pi^{+}\pi^{0}\pi^{0}\pi^{0},
\]
\begin{equation}
\tau^{+}\rightarrow W^{+}\bar{\nu}_{\tau} \rightarrow \tilde{\rho}^{+}
\bar{\nu}_{\tau}\rightarrow \bar{\nu}_{\tau}\pi^{+}\pi^{-}\pi^{+}\pi^{0}.
\label{tau}
\end{equation} 
The relations between processes (\ref{ee}) and (\ref{tau}) can be written as
\begin{equation}
\Gamma(\tilde{\rho}^{+}\rightarrow \pi^{+}\pi^{-}\pi^{+}\pi^{0})
=\frac{1}{2}\Gamma(\tilde{\rho}^{0}\rightarrow \pi^{+}\pi^{-}\pi^{+}\pi^{-})
+\Gamma(\tilde{\rho}^{0}\rightarrow \pi^{+}\pi^{-}\pi^{0}\pi^{0}),
\label{channel1}
\end{equation}
\begin{equation}
\Gamma(\tilde{\rho}^{+}\rightarrow \pi^{+}\pi^{0}\pi^{0}\pi^{0})=
\frac{1}{2}\Gamma(\tilde{\rho}^{0}\rightarrow \pi^{+}\pi^{-}\pi^{+}\pi^{-}).
\label{channel2}
\end{equation} 
The prediction  for $\Gamma(\tau^{+}\rightarrow\bar{\nu}_{\tau}
X^{+})$
can thus be obtained from $e^{+}e^{-}\rightarrow 4\pi$ data and isospin
invariance. Such a procedure was successfully 
applied in \cite{Binp1} where high statistics $e^{+}e^{-}$ data from 
the CMD-2 detector collected at center of mass energies ($Q$) from $1.05$ 
to $1.38$ GeV were used \cite{Binp2}. 
Since $e^{+}e^{-}$ experiments are performed at fixed (at a time) $Q$,
the integrated decay rates of  $\tilde{\rho}^{0}$ to the $4\pi$ 
are well measured as a function of $Q$.  
Thus, the natural way of the $\tau$ decay generation is to generate first     
the mass of the $4\pi$ system in accordance with the experimental 
distribution. Then, for the fixed $Q^{2}$, the  
$\tilde{\rho}^{0}\rightarrow4\pi$ decay is generated.
This is why in Ref. \cite{Binp1} the  $d\Gamma/dQ^{2}$ 
distribution is generated independently of the differential
distribution within the $4\pi$ system.  
The program is written and optimized to get maximum possible 
information from the experimental data.

The approach of {\tt TAUOLA}  is somewhat different. 
A matrix element well isolated in a
program module is separated into the hadronic and leptonic current.
All physical  assumptions on hadronic interactions are located in the
hadronic current which features
intermediate state resonances as well as other properties of the hard process.
The phase space density generation and in particular the appropriate 
jacobians for mapping  
 random numbers to phase space coordinates, 
which are defined independently from the particular process, 
are also  calculated in the separate part of the code. 
Such an approach provides flexibility
in studying particular choices of hadronic currents.
In addition, the $Q^2$ distribution is not an input, but originates from 
the (partial) Monte Carlo integration
of the matrix element over phase space.  

The aim of this paper is to use the model of \cite{Binp1}
in the approach of {\tt TAUOLA} which is more natural for comparison with 
other theoretical calculations/models.

\section{General formalism for semileptonic decays }
\vskip 0.3 cm

The matrix element used in {\tt TAUOLA} for the semileptonic decay $\tau(P,s)\rightarrow
\nu_{\tau}(N)X$ is 
written in the form:
\begin{equation}
{\cal M}=\frac{G}{\sqrt{2}}\bar{u}(N)\gamma^{\mu}(v+a\gamma_{5})u(P)J_{\mu}
\end{equation}
where $J_{\mu}\equiv<X|V_{\mu}-A_{\mu}|0>$ denotes the matrix element of the 
$V -A$ current, relevant for the specific final state X. In general,
the current $J_{\mu}$ depends on the momenta of all hadrons. $N$ and $P$ 
denote the four-momenta of the $\nu_{\tau}$ and $\tau$ respectively. 
The squared matrix element for the decay of $\tau$ with mass $M$ and
 spin $s$ reads:
\[
|{\cal M}|^{2}= G^{2}\frac{v^{2}+a^{2}}{2}( \omega + H_{\mu}s^{\mu} ),
\]
\[
\omega=P^{\mu}(\Pi_{\mu}-\gamma_{va}\Pi_{\mu}^{5}),
\]
\begin{equation}
H_{\mu}=\frac{1}{M}(M^{2}\delta^{\nu}_{\mu}-P_{\mu}P^{\nu})(\Pi_{\nu}^{5}-
\gamma_{va}\Pi_{\nu})
\end{equation}
with
\[
\Pi_{\mu}=2[(J^{*}\cdot N)J_{\mu}+(J\cdot N)J_{\mu}^{*}-(J^{*}\cdot J)N_{\mu}],
\]
\[
\Pi^{5\mu}=2~ {\rm Im} ~\epsilon^{\mu\nu\rho\sigma}J^{*}_{\nu}J_{\rho}N_{\sigma},
\]
\begin{equation}
\gamma_{va}=-\frac{2va}{v^{2}+a^{2}}
\end{equation}
($\gamma_{va}=1$ in the Standard Model). If a more general coupling 
$v+a\gamma_{5}$ for the $\tau$ current and $\nu_{\tau}$ mass $m_{\nu} \neq 0$ 
are expected to be used, one has
to add the following terms to $\omega$ and $H_{\mu}$:
\[
\hat{\omega}=2\frac{v^{2}-a^{2}}{v^{2}+a^{2}}m_{\nu}M(J^{*} \cdot J),
\]
\begin{equation}
\hat{H}^{\mu}=-2\frac{v^{2}-a^{2}}{v^{2}+a^{2}}m_{\nu}~ {\rm Im}~\epsilon^{\mu\nu\rho\sigma}
J_{\nu}^{*}J_{\rho}P_{\sigma}.
\end{equation}
To obtain the polarimeter vector $h$ in the $\tau$ rest frame, it is 
sufficient to calculate the space components of 
$h_{\mu}=(H_{\mu}+\hat{H_{\mu}})/(\omega+\hat{\omega})$ and set $h_{0}=0$. 
The differential partial width
for the  channel under consideration reads:
\begin{equation}
d\Gamma_{X}=G^{2}\frac{v^{2}+a^{2}}{4M}d{\rm Lips}(P;q_{i},N)(\omega+\hat{\omega}+
(H_{\mu}+\hat{H_{\mu}})s^{\mu}).
\end{equation}
The phase space distribution for the final state with four mesons plus 
neutrino is given by the following expression where a compact notation with
 $q_{5}=N$ and $q_{i}^{2}=m^{2}_{i}$ is used, 
\[
 d{\rm Lips}(P;q_{1},q_{2},q_{3},q_{4},q_{5})=\frac{1}{2^{23}\pi^{11}}~dQ^{2}~
dQ_{3}^{2}~dQ_{2}^{2}~~~~~\times
\]
\begin{equation}
d\Omega_{5}\frac{\sqrt{\lambda(M^{2},Q^{2},m_{5}^{2})}}{M^{2}}
d\Omega_{4}\frac{\sqrt{\lambda(Q^{2},Q_{3}^{2},m_{4}^{2})}}{Q^{2}}
d\Omega_{3}\frac{\sqrt{\lambda(Q_{3}^{2},Q_{2}^{2},m_{3}^{2})}}{Q_{3}^{2}}
d\Omega_{2}\frac{\sqrt{\lambda(Q_{2}^{2},m_{2}^{2},m_{1}^{2})}}{Q_{2}^{2}}
\end{equation}
where
\[
Q^{2}=(q_{1}+q_{2}+q_{3}+q_{4})^{2},~~~~~~Q_{3}^{2}=
(q_{1}+q_{2}+q_{3})^{2},~~~~~~
Q^{2}_{2}=(q_{1}+q_{2})^{2},
\]
\[
Q_{min}=m_{1}+m_{2}+m_{3}+m_{4}, ~~~~~~Q_{max}=M-m_{5},
\]
\[
Q_{3,min}=m_{1}+m_{2}+m_{3}, ~~~~~~~~~~~Q_{3,max}=Q-m_{4},
\]
\begin{equation}
Q_{2,min}=m_{1}+m_{2}, ~~~~~~~~~~~~~~~~~Q_{2,max}=Q_{3}-m_{3}.
\end{equation}
Here $d\Omega_{5}=d\cos\theta_{5}d\pi_{5}$ is the solid angle element 
of the momentum of $\nu_{\tau}$ in the rest frame of $\tau(P)$, 
$d\Omega_{4}=d\cos\theta_{4}d\pi_{4}$ is the solid angle element of 
$\vec{q}_{4}$ in the rest frame of 
$q_{1}^{\mu}+q_{2}^{\mu}+q_{3}^{\mu}+q_{4}^{\mu}$, 
$d\Omega_{3}=d\cos\theta_{3}d\pi_{3}$ is the solid angle element of 
$\vec{q}_{3}$ in the $q_{1}^{\mu}+q_{2}^{\mu}+q_{3}^{\mu}$ rest frame, and 
finally, $d\Omega_{2}=d\cos\theta_{2}d\pi_{2}$ is the solid angle element of 
$\vec{q}_{2}$ in the $q_{1}^{\mu}+q_{2}^{\mu}$ rest frame.\\
These formula if used directly, are inefficient for a Monte Carlo algorithm 
if sharp peaks due 
to resonances in the intermediate states
are present. We refer to the {\tt TAUOLA} documentation
\cite{tauola:1990,tauola:1993}  for details of the algorithm actually in use. 
For the present paper it is enough to note that those
changes affect the program efficiency, but the actual density of 
the phase space remains intact. No approximations are introduced.

\vskip 0.4 cm

\section{Hadronic current for 4$\pi$ system}
\vskip 0.3 cm 

The model of Ref.~\cite{Binp1} is based on the assumption that the 
$a_{1}(1260)\pi$ and $\omega\pi$ intermediate states (which well describe
the experiments on $e^+e^- \to 4\pi$ \cite{Binp2}), are dominant 
in the amplitudes $\tau^{+}\rightarrow\bar{\nu}_{\tau}
\tilde{\rho}^{+}\rightarrow\bar{\nu}_{\tau}
(4\pi)^{+}$. In Ref.~\cite{Binp1} it was shown that various
two- and three-pion invariant mass distributions predicted by the
model well describe experimental observations of CLEO~\cite{CLEO2000} and
ALEPH~\cite{aleph}.   
We include into consideration two  most important channels 
of the $a_{1}\rightarrow3\pi$ decay 
($a_{1}\rightarrow \rho\pi \rightarrow 3\pi$ and
$a_{1}\rightarrow \sigma\pi \rightarrow 3\pi$) as well as the 
$\omega\rightarrow\rho\pi \rightarrow 3\pi$ channel. 
Then for the process
$\tau^{+}\rightarrow\bar{\nu}_{\tau}\pi^{+}\pi^{0}\pi^{0}\pi^{0}$ 
the current $J^{\mu}$ reads 
\begin{equation}
J^{\mu}=J^{\mu}_{a_{1}\rightarrow\rho\pi}+J^{\mu}_{a_{1}\rightarrow\sigma\pi}.
\end{equation} 
For the process $\tau^{+}\rightarrow\bar{\nu}_{\tau}\pi^{+}
\pi^{-}\pi^{+}\pi^{0}$, where the $\omega$ meson also contributes, it is
\begin{equation}
J^{\mu}=J^{\mu}_{a_{1}\rightarrow\rho\pi}+J^{\mu}_{a_{1}\rightarrow\sigma\pi}
+J^{\mu}_{\omega\rightarrow\rho\pi}
\end{equation}
where in the following we neglect the interference between the $\omega$ and 
$a_{1}$ currents.
The $\bar{\nu}_{\tau}\pi^+\pi^-\pi^+\pi^0$ final states produced with the two 
currents ($a_1(1260)\pi$ and $\omega\pi$) 
are effectively treated as distinct tau decay modes.

\subsection{$\tau^{+}\rightarrow\bar{\nu}_{\tau}\pi^{+}\pi^{0}
\pi^{0}\pi^{0}$ decay channel}
\vskip 0.3 cm
 
For the $\tau^{+}\rightarrow\bar{\nu}_{\tau}\pi^{+}(q_{1})\pi^{0}(q_{2})
\pi^{0}(q_{3})\pi^{0}(q_{4})$  channel the current which includes  possible 
Feynman diagrams 
\footnote{~Possible Feynman
 diagrams for $\tau^{+}$
decays into $4\pi$ via the $a_{1}\pi$ and the $\omega\pi$ 
intermediate states 
are shown in Appendix A. }, can be written in the following way:
\[
J^{\mu}_{a_1\rightarrow\rho\pi}=G_{\pi^{+}\pi^{0}\pi^{0}\pi^{0}}
(Q^{2})[t_{1}^{\mu}(q_{2},q_{3},q_{1},q_{4})+
t_{1}^{\mu}(q_{2},q_{4},q_{1},q_{3})+t_{1}^{\mu}(q_{3},q_{2},q_{1},q_{4})
\]
\begin{equation}
~~~~~~~~~~~+t_{1}^{\mu}(q_{3},q_{4},q_{1},q_{2})+t_{1}^{\mu}
(q_{4},q_{2},q_{1},q_{3})+
t_{1}^{\mu}(q_{4},q_{3},q_{1},q_{2})],
\end{equation}
\[
J^{\mu}_{a_1\rightarrow\sigma\pi}=G_{\pi^{+}\pi^{0}\pi^{0}\pi^{0}}
(Q^{2})[t_{2}^{\mu}(q_{2},q_{1},q_{3},q_{4})+
t_{2}^{\mu}(q_{3},q_{1},q_{2},q_{4})+t_{2}^{\mu}(q_{4},q_{1},q_{3},q_{2})
\]
\begin{equation}
~~~~~~~~~~~-t_{2}^{\mu}(q_{1},q_{2},q_{3},q_{4})-t_{2}^{\mu}
(q_{1},q_{3},q_{2},q_{4})-
t_{2}^{\mu}(q_{1},q_{4},q_{3},q_{2})].
\end{equation}
Four-vectors $t_{1}^{\mu}$ and $t_{2}^{\mu}$ have the following forms, where
 $Q$ 
denotes $Q=q_{1}+q_{2}+q_{3}+q_{4}$: 
\[
t_{1}^{\mu}(q_{1},q_{2},q_{3},q_{4})=\frac{F_{a_1}^{2}(Q-q_{1})}
{D_{a_1}(Q-q_{1})D_{\rho}(q_{3}+q_{4})}~~~~~~\times
\]
\[
\{ Q\cdot(Q-q_{1})[q_{4}^{\mu}(Q-q_{1})\cdot q_{3}-q_{3}^{\mu}
(Q-q_{1})\cdot q_{4}] 
\]
\begin{equation}
+(Q^{\mu}-q_{1}^{\mu})[(Q\cdot q_{4})(q_{1}\cdot q_{3})-(
Q\cdot q_{3})(q_{4}\cdot q_{1})] \}
\end{equation}
\[
t_{2}^{\mu}(q_{1},q_{2},q_{3},q_{4})=\frac{{\it z}F_{a_1}^{2}(Q-q_{1})}
{D_{a_1}(Q-q_{1})D_{\sigma}(q_{3}+q_{4})}~~~~~~\times
\]
\begin{equation}
\{q_{2}^{\mu}Q\cdot (Q-q_{1})(Q-q_{1})^{2}+(q_{1}^{\mu}-Q^{\mu})
[(Q\cdot q_{2})(Q-q_{1})^{2}] 
\}.
\end{equation}
Here $1/D_{a_1}(q)$, $1/D_{\rho}(q)$ and $1/D_{\sigma}(q)$ are propagators 
of the $a_{1}$,
$\rho$ and $\sigma$ mesons, $F_{a_1}(q)$ is the form factor and ${\it z}$
is the dimensionless complex constant characterizing the relative 
fraction of the $\sigma\pi$ intermediate state in the $a_1(1260)$ decay. 
$G_{\pi^{+}\pi^{0}\pi^{0}\pi^{0}}$ is some function depending on 
$Q^{2}$ which we find by fitting the $4\pi$ invariant 
mass distribution~\footnote{~For details see chapter
{\bf The $G(Q^{2})$ functions}.}.
 As a $F_{a_1}$ form factor, we used the function from \cite{Binp1,Binp2},
$F(q)=(1+m^{2}_{a_1}/ \Lambda^{2})/(1+q^{2}/ \Lambda^{2})$ with $\Lambda\sim1$ 
GeV. 

The form of the propagators is very important for analyzing the data. We 
represent the function $D(q)$ in the form used in \cite{Binp1,Binp2} 
\begin{equation}
D(q)=q^{2}-M^{2}+iM\Gamma\frac{g(q^{2})}{g(M^{2})},
\end{equation}
where $M$ and $\Gamma$ are the mass and width of the corresponding particle, 
and the function $g(s)$ describes the dependence of the width on 
virtuality. In the case of the $\rho$ meson the function $g_{\rho}(s)$ reads:
\begin{equation}
g_{\rho}(s)=s^{-1/2}(s-4m^{2})^{3/2},
\end{equation}
while for the  $\sigma$ meson it is:
\begin{equation}
g_{\sigma}(s)=(s-4m^{2}/s)^{1/2},
\end{equation}
where $m$ is the pion mass. 

The function $g_{a_{1}}$ in the $a_{1}$ propagator 
has the form:

\[
g_{a_1}(s)= F_{a_1}^2(q) \int \Biggl\{  \left|
 \frac{\varepsilon_{2} {\bf p}_1-\varepsilon_1 {\bf p}_2}{D_{\rho}(p_1+p_2)}+
 \frac{\varepsilon_2 {\bf p}_3- \varepsilon_3{\bf p}_2}{D_{\rho}(p_2+p_3)}
+\frac{z\sqrt{s}{\bf p}_2}{D_{\sigma}(p_1+p_3)}\right|^2 
\]

\[
+\frac{|z|^2s}{3!}\left |
\frac{{\bf p}_1}{D_{\sigma}(p_2+p_3)}+
\frac{{\bf p}_2}{D_{\sigma}(p_1+p_3)}+\frac{{\bf p}_3}
{D_{\sigma}(p_1+p_2)}\right|^2  ~ \Biggr\}
 \times
\]
\begin{equation}
\times \frac{d{\bf p}_1\,d{\bf p}_2\,d{\bf p}_3~\delta^{(4)}(p_1+p_2+p_3-q)}
 {2\varepsilon_1 2\varepsilon_2 2\varepsilon_3(2\pi)^5}
\end{equation}
where $q^0=\sqrt s$, ${\bf q}=0$ and 
$p_{i}=(\varepsilon_{i},{\bf p}_{i})$
are the pion momenta in the rest frame of the $\tilde{\rho}$ 
(the center of mass frame of the $4\pi$ system).
 The first term corresponds
to the $a_1\to\pi^+\pi^-\pi^0$ decay while the second one to the 
$a_1\to 3\pi^0$ decay.

\subsection{$\tau^{+}\rightarrow\bar{\nu}_{\tau}\pi^{+}
\pi^{-}\pi^{+}\pi^{0}$ decay channel}
For the $\tau^{+}\rightarrow\bar{\nu}_{\tau}\pi^{+}(q_{1})\pi^{-}(q_{2})
\pi^{+}(q_{3})\pi^{0}(q_{4})$  channel the current which includes 
the contribution from the $\omega$ meson
intermediate state can be written in the following way: 
\[
J^{\mu}_{a_1\rightarrow\rho\pi}=G_{\pi^{+}\pi^{-}\pi^{+}\pi^{0}}(Q^{2})
[t_{1}^{\mu}(q_{1},q_{2},q_{3},q_{4})+
t_{1}^{\mu}(q_{3},q_{2},q_{1},q_{4})+t_{1}^{\mu}(q_{1},q_{3},q_{2},q_{4})
\]
\begin{equation}
~~~~~~~~~~~+t_{1}^{\mu}(q_{3},q_{1},q_{2},q_{4})+t_{1}^{\mu}
(q_{4},q_{3},q_{1},q_{2})+
t_{1}^{\mu}(q_{4},q_{1},q_{3},q_{2})],
\end{equation}
\[
J^{\mu}_{a_1\rightarrow\sigma\pi}=G_{\pi^{+}\pi^{-}\pi^{+}\pi^{0}}(Q^{2})
[t_{2}^{\mu}(q_{4},q_{3},q_{1},q_{2})+
t_{2}^{\mu}(q_{4},q_{1},q_{3},q_{2})
\]
\begin{equation}
~~~~~~~~~~~-t_{2}^{\mu}(q_{1},q_{4},q_{3},q_{2})-t_{2}^{\mu}
(q_{3},q_{4},q_{1},q_{2})],
\end{equation}
\[
J^{\mu}_{\omega\rightarrow\rho\pi}=G^{\omega}_{\pi^{+}\pi^{-}\pi^{+}
\pi^{0}}(Q^{2})
[t_{3}^{\mu}(q_{1},q_{2},q_{3},q_{4})+
t_{3}^{\mu}(q_{3},q_{2},q_{1},q_{4})-t_{3}^{\mu}(q_{1},q_{3},q_{2},q_{4})
\]
\begin{equation}
~~~~~~~~~~~-t_{3}^{\mu}(q_{3},q_{1},q_{2},q_{4})
-t_{3}^{\mu}(q_{1},q_{4},q_{3},q_{2})-t_{3}^{\mu}(q_{3},q_{4},q_{1},q_{2})].
\end{equation}
Here four-vectors $t_{1}^{\mu}$, $t_{2}^{\mu}$  are the same as in the previous
 case and $t_{3}^{\mu}$ reads: 
\[
t_{3}^{\mu}(q_{1},q_{2},q_{3},q_{4})=\frac{F_{\omega}^{2}(Q-q_{1})}
{D_{\omega}(Q-q_{1})D_{\rho}(q_{3}+q_{4})}~~~~~~\times
\]
\[
\{q_{2}^{\mu}[(Q\cdot q_{3})(q_{1}\cdot q_{4})-(Q\cdot q_{4})(q_{1}\cdot 
q_{3})]-Q\cdot q_{2}[q_{3}^{\mu}(q_{1}\cdot q_{4})-q_{4}^{\mu}(q_{1}\cdot 
q_{3})]
\]
\begin{equation}
+(q_{1}\cdot q_{2})[q_{3}^{\mu}(Q\cdot q_{4})-q_{4}^{\mu}(Q\cdot q_{3})] 
\},
\end{equation}
where $F_{\omega}(q)$ and $D_{\omega}(q)$ are the form factor and propagator 
for the $\omega$. Because of the small width of the $\omega$ we set 
$F_{\omega}(q)=1$ and $g_{\omega}(s)=1$. $G_{\pi^{+}\pi^{-}\pi^{+}\pi^{0}}$  
and $G^{\omega}_{\pi^{+}\pi^{-}\pi^{+}\pi^{0}}$  are functions of $Q^{2}$.

\vskip 0.4 cm

\section{Parameters used in the model}

\begin{table}[h!]
\newcommand{\lstrut}{{$\strut\atop\strut$}}
  \caption {\em The masses and widths for the intermediate states.
\label{masses}}
\vspace{2mm}
\begin{center}
\begin{tabular}{|c|c|c|c| } \hline \hline 
 Intermediate          &  Mass, GeV    & Width, GeV        \\ 
 state                 &       &              \\ \hline \hline
 $\rho(770)$                & 0.7761         & 0.1445        \\ \hline
 $a_{1}(1260)$              & 1.23           & 0.45          \\ \hline
 $\omega(782)$              & 0.782          & 0.00841       \\ \hline
 $\sigma$                   & 0.8            & 0.8           \\ \hline \hline
\end{tabular}
\end{center}
\end{table}
The  constants, widths and other parameters used in our 
numerical results were mainly taken from \cite{parameters},
for the $\rho$ meson recent CMD-2 results were used \cite{rho}. 
Some of them  are collected in 
Table~\ref{masses},
The $m_{\nu}=0$ was assumed
 and  {\tt PHOTOS} \cite{photos1,photos2} 
for QED radiative corrections was switched off.
We used a $\tau$ lepton mass of $m_{\tau}$=1.777 GeV
and the physical masses for the pions in the phase space, {\it i.e.}
$m_{\pi^{\pm}}=0.13957018$ GeV, $m_{\pi^{0}}=0.1349766$ GeV.
Note that because of CVC the masses of $\pi^{\pm}$ and $\pi^{0}$ were set 
equal $m_{\pi^{0}}=m_{\pi^{\pm}}=0.139570$ in the hadronic current. 
 The parameter $\Lambda=1.2$ GeV for the
$F_{a_1}$ form factor was taken. The dimensionless complex constant ${\it z}$
for $t_{2}^{\mu}(q_{1},q_{2},q_{3},q_{4})$ and for $g_{a_1}(s)$
was set to $(1.269,0.591)$ which means that the ratio of
contributions of the  $\sigma$ to $\rho$
intermediate states is 0.3 accordingly to Ref. \cite{hep-ex/9902022}.

%

\subsection{The $G(Q^2)$ functions}
New channels in {\tt TAUOLA} are based on the same matrix elements for 
the $\tilde{\rho}^{+}$ decay as in \cite{Binp1}. The difference is in $Q^{2}$ 
generation. To complete the definition of the hadronic current,
the appropriate choice of the three functions $G(Q^{2})$  
was needed.
We fit the functions
$G_{\pi^{+}\pi^{0}\pi^{0}\pi^{0}}(Q^{2})$, $G_{\pi^{+}\pi^{-}\pi^{+}\pi^{0}}
(Q^{2})$, and $G^\omega_{\pi^{+}\pi^{-}\pi^{+}\pi^{0}}(Q^{2})$ 
in such a way, that {\tt TAUOLA} reproduces
predictions from  the calculation \cite{Binp1} for the single variable 
$d \Gamma/d Q^2$ distribution. Samples  of 5 000 000 events were used. 
The table of numerical values for the fitted functions is 
given in  Appendix B. The functions 
are well determined by the $e^+e^-$ data  from 0.9 to 1.7 GeV,
for lower energies the uncertainties are larger since the  measured 
cross section is  small, and we rely on assumptions made in Ref.~\cite{Binp1}. 

Finally, we have normalized the functions $G(Q^2)$ in such a way that our program reproduces 
$\tau$ decay rates of Ref. \cite{parameters}, see results in  
Table~\ref{decay_rates}.
\begin{table}[!h]
\newcommand{\lstrut}{{$\strut\atop\strut$}}
  \caption {\em Numerical results for the integrated decay rates of our model }
\label{decay_rates}
\vspace{2mm}
\begin{center}
\begin{tabular}{|c|c|} \hline \hline                
  Channel                         & Decay rate, $GeV$ \\\hline \hline
$\Gamma(\tau^{+}\rightarrow\bar{\nu}_{\tau}\pi^{0}\pi^{0}\pi^{0}\pi^{+})$ &
$2.4462 *10^{-14}$  \\ \hline
$\Gamma(\tau^{+}\rightarrow\bar{\nu}_{\tau}\pi^{+}\pi^{+}\pi^{-}\pi^{0})$ &
$9.5130 *10^{-14}$  \\ \hline \hline
\end{tabular}
\end{center}
\end{table}

\vskip 0.4 cm
\section{ Numerical tests}

\subsection{Technical test}
While developing the Monte Carlo algorithm, it is important to 
perform  numerous technical tests of the Monte Carlo code.
Tests of the {\tt TAUOLA} Monte Carlo library consisting of
the comparison of numerical results obtained from 
the program and independent semi-analytical calculations are described and 
listed in Ref.\cite{tauola:1993}.
At that time the agreement between the Monte Carlo results and some 
analytical calculations 
was pushed to the level of few permille. Recently, thanks to much faster 
computers, checks at the 0.1 \% level were in some cases redone. In 
those tests different assumptions about matrix
elements were used, {\it e.g.} masses of $\pi^\pm$ and $\pi^0$ were set equal 
or even zero. The actual choice of the current was 
also modified. Parameters of the presamples were varied and results were 
checked to be independent of that.
Satisfactory agreement was always found. This gives us confidence in 
the technical side of the algorithm.

\subsection{General test}

We  have compared various possible invariant mass distributions
constructed from the momenta of the $\tau$ decay products
in the Novosibirsk model coded in {\tt TAUOLA} on one side 
to the code used in Ref. \cite{Binp1} on the other side.
The samples used in the plots were of the size of the experimental
data used as an input to the model, {\it i.e.} 30 000 events.
In each case agreement was sufficiently good, for details see~\cite{Tomek}.  
In this way we could convince ourselves that the adaptation was 
really correct.

\section{Comparison with other parameterizations }
\vskip 0.3 cm

\begin{figure}[!ht]
\setlength{\unitlength}{0.1mm}
\begin{picture}(1600,800)
\put( 375,750){\makebox(0,0)[b]{\large }}
\put(1225,750){\makebox(0,0)[b]{\large }}
\put(-20, -1){\makebox(0,0)[lb]{\epsfig{file=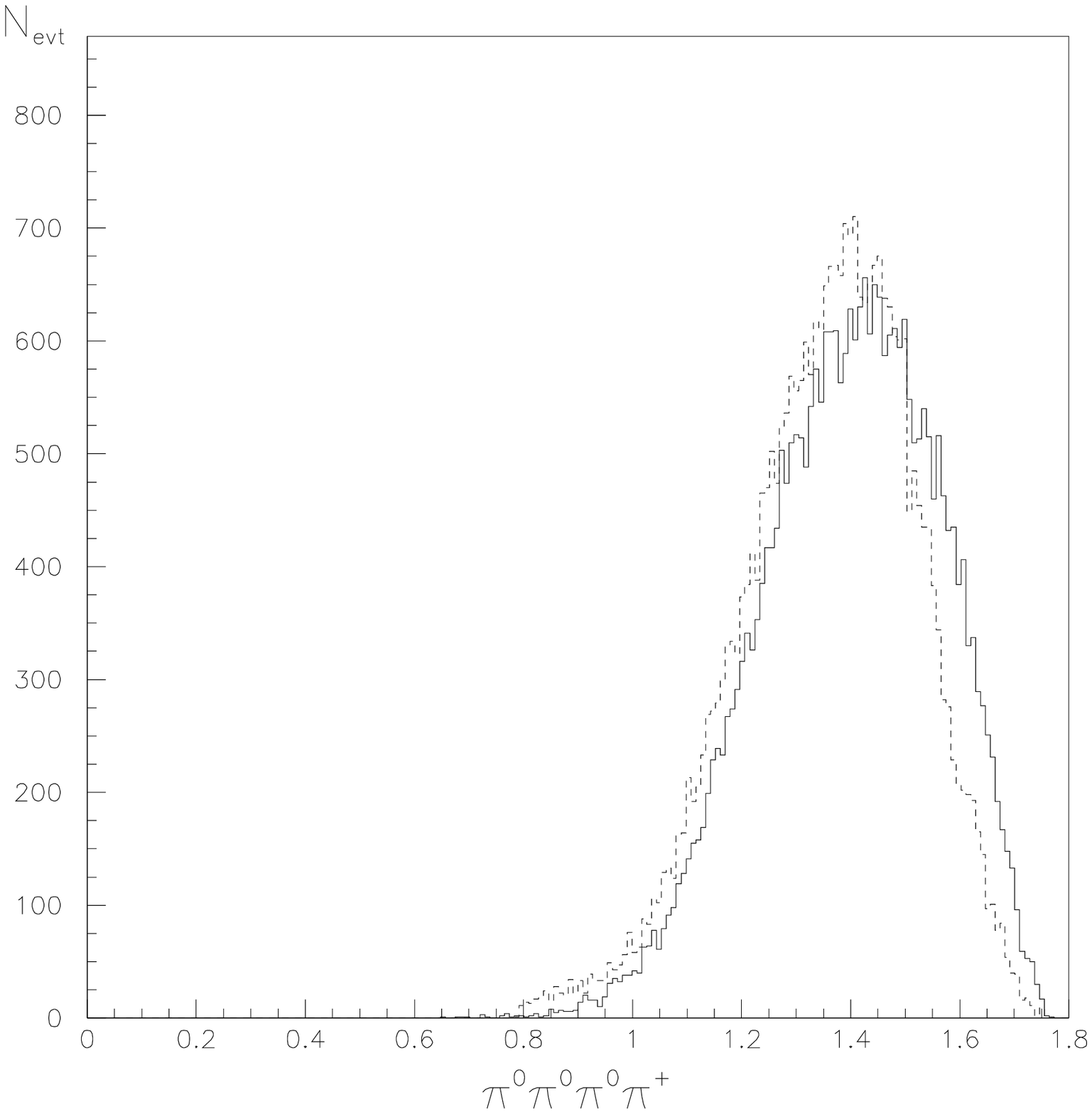,width=80mm,height=80mm}}}
\put(700, -1){\makebox(0,0)[lb]{\epsfig{file=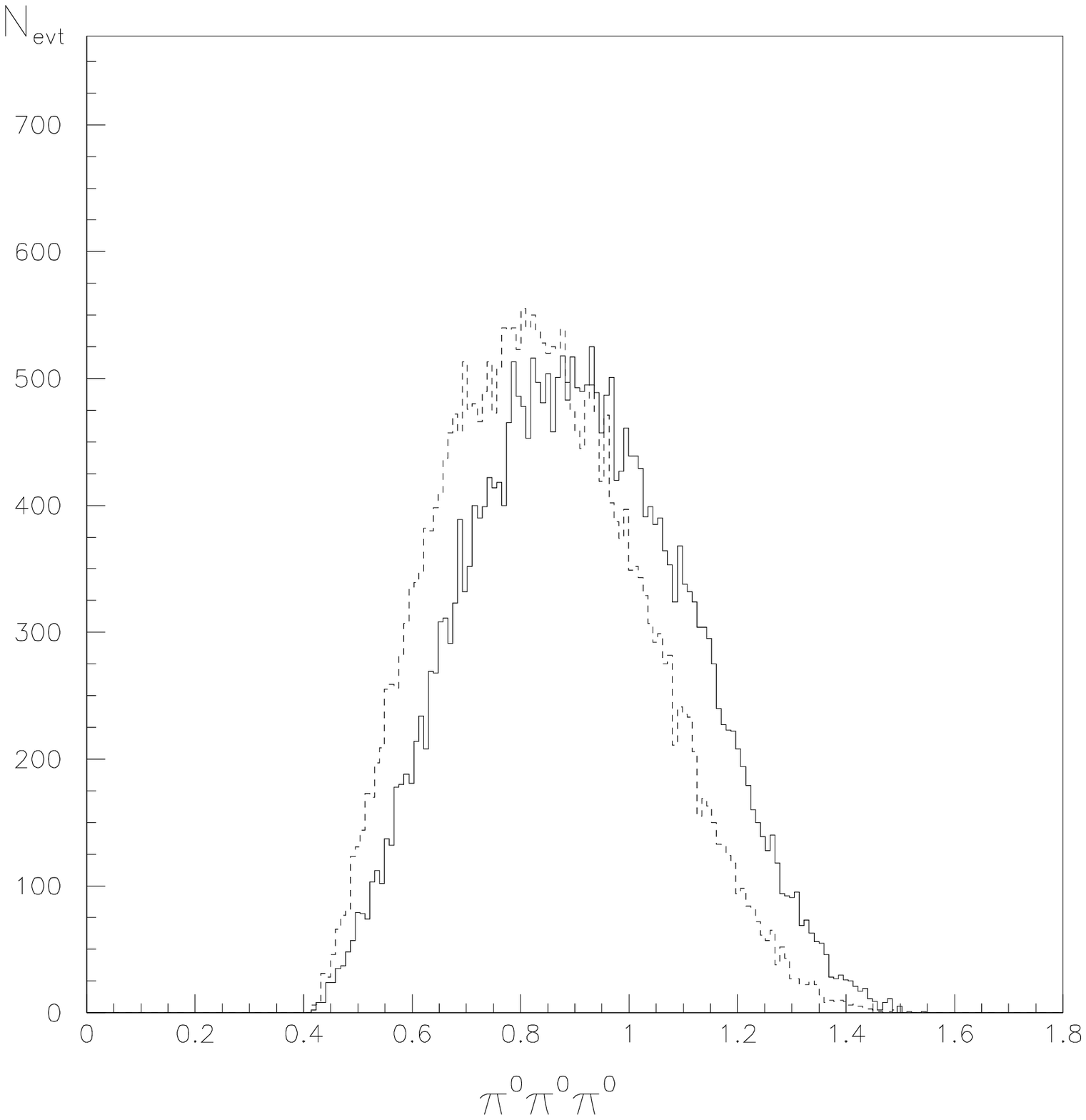,width=80mm,height=80mm}}}
\end{picture}
\caption
{\it The $\bar{\nu}_{\tau}\pi^{0}\pi^{0}\pi^{0}\pi^{+}$ channel. The  
left-hand side plots the $\pi^{0}\pi^{0}\pi^{0}\pi^{+}$ invariant
mass distribution and the right-hand side is the $\pi^{0}\pi^{0}\pi^{0}$ 
invariant mass distribution. The continuous and dotted lines correspond to the
old CLEO  and new Novosibirsk current respectively.}
\label{3pi0pi1}
\end{figure}
\begin{figure}[!ht]
\setlength{\unitlength}{0.1mm}
\begin{picture}(1600,800)
\put( 375,750){\makebox(0,0)[b]{\large }}
\put(1225,750){\makebox(0,0)[b]{\large }}
\put(-20, -1){\makebox(0,0)[lb]{\epsfig{file=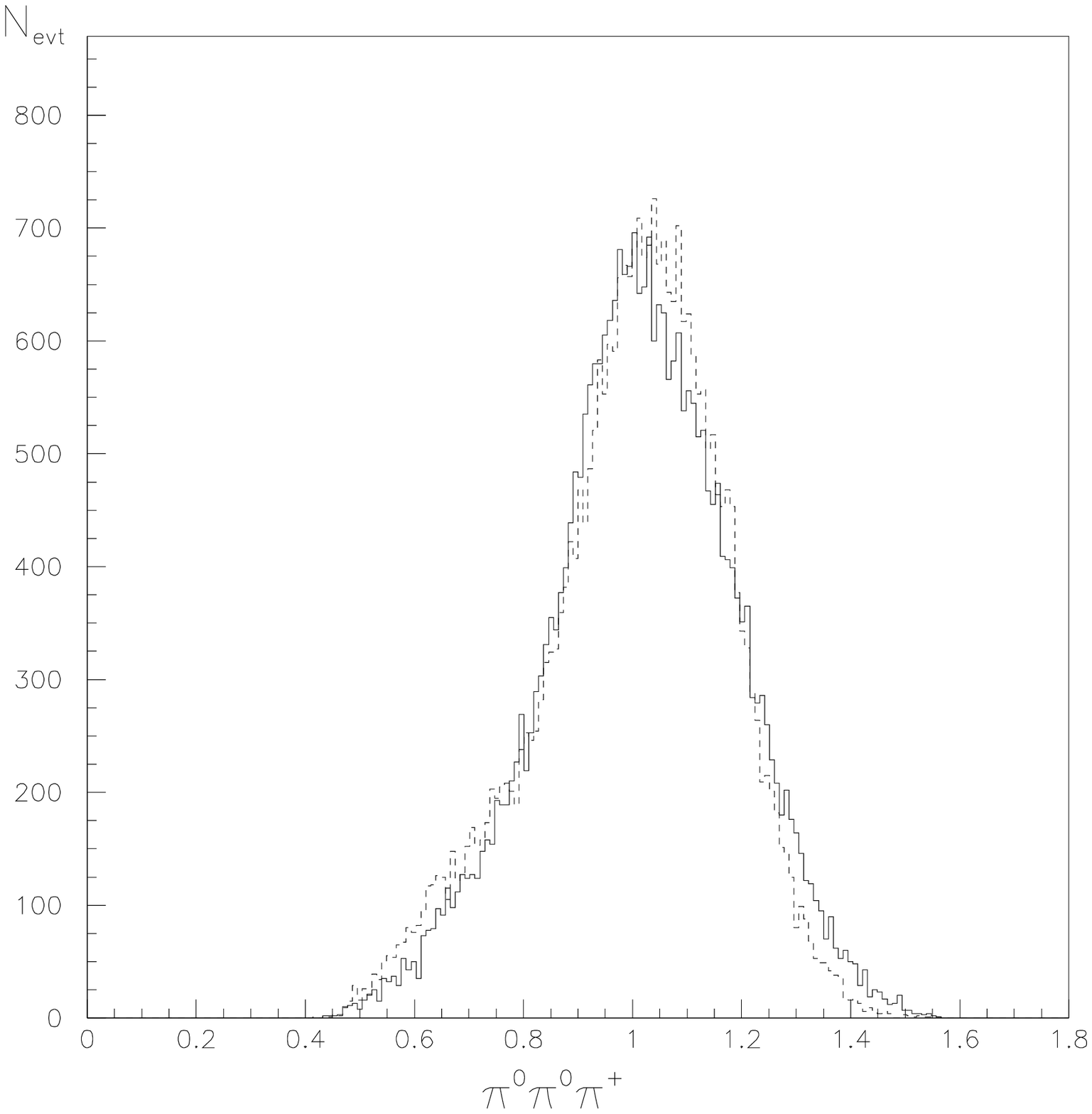,width=80mm,height=80mm}}}
\put(700, -1){\makebox(0,0)[lb]{\epsfig{file=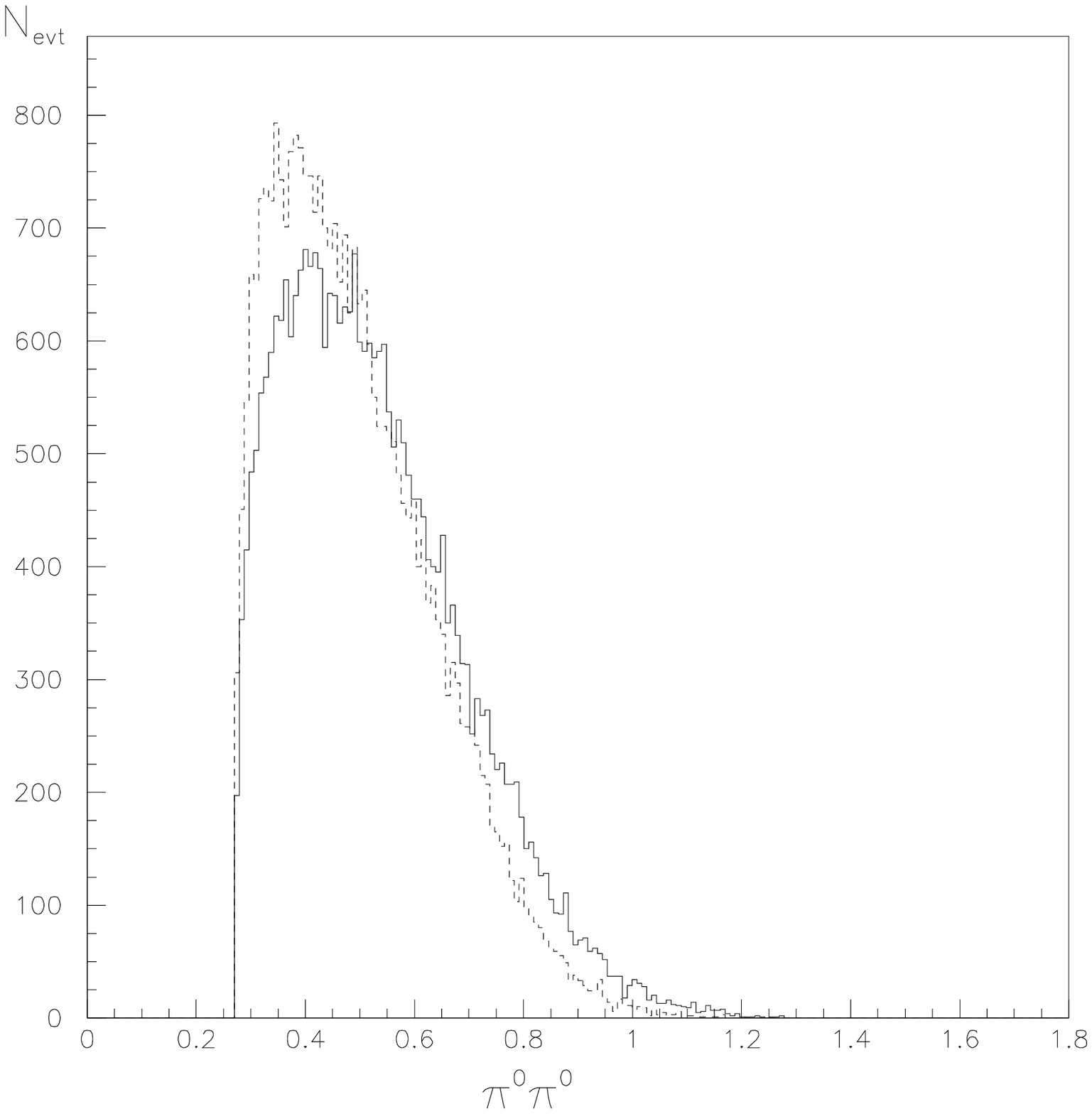,width=80mm,height=80mm}}}
\end{picture}
\caption 
{\it The $\bar{\nu}_{\tau}\pi^{0}\pi^{0}\pi^{0}\pi^{+}$ channel. The left-hand 
side plots the $\pi^{0}\pi^{0}\pi^{+}$ invariant
mass distribution and the right-hand side is the $\pi^{0}\pi^{0}$ 
invariant mass distribution. The continuous and dotted lines
correspond to the old CLEO and new Novosibirsk current respectively. }

\label{3pi0pi2}
\end{figure}
\begin{figure}[!ht]
\setlength{\unitlength}{0.1mm}
\begin{picture}(1600,800)
\put( 375,750){\makebox(0,0)[b]{\large }}
\put(1225,750){\makebox(0,0)[b]{\large }}
\put(400, -1){\makebox(0,0)[lb]{\epsfig{file=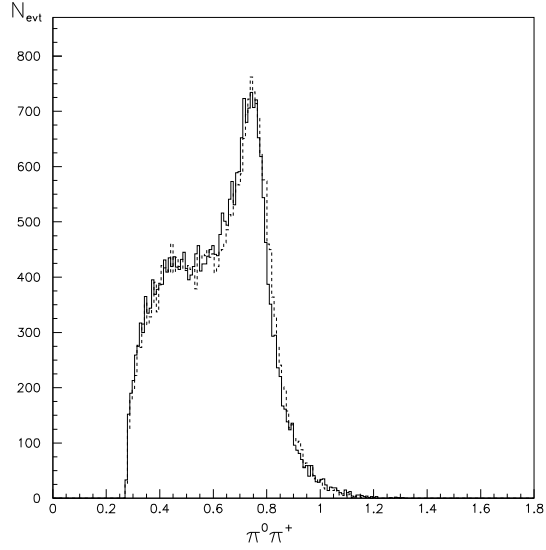,width=80mm,height=80mm}}}
\end{picture}
\caption 
{\it The $\pi^{0}\pi^{+}$ invariant mass distribution 
for the $\bar{\nu}_{\tau}\pi^{0}\pi^{0}\pi^{0}\pi^{+}$ channel.
The continuous and dotted lines correspond to the old CLEO  and
new Novosibirsk current respectively.}

\label{3pi0pi3}
\end{figure}
\begin{figure}[!ht]
\setlength{\unitlength}{0.1mm}
\begin{picture}(1600,800)
\put( 375,750){\makebox(0,0)[b]{\large }}
\put(1225,750){\makebox(0,0)[b]{\large }}
\put(-20, -1){\makebox(0,0)[lb]{\epsfig{file=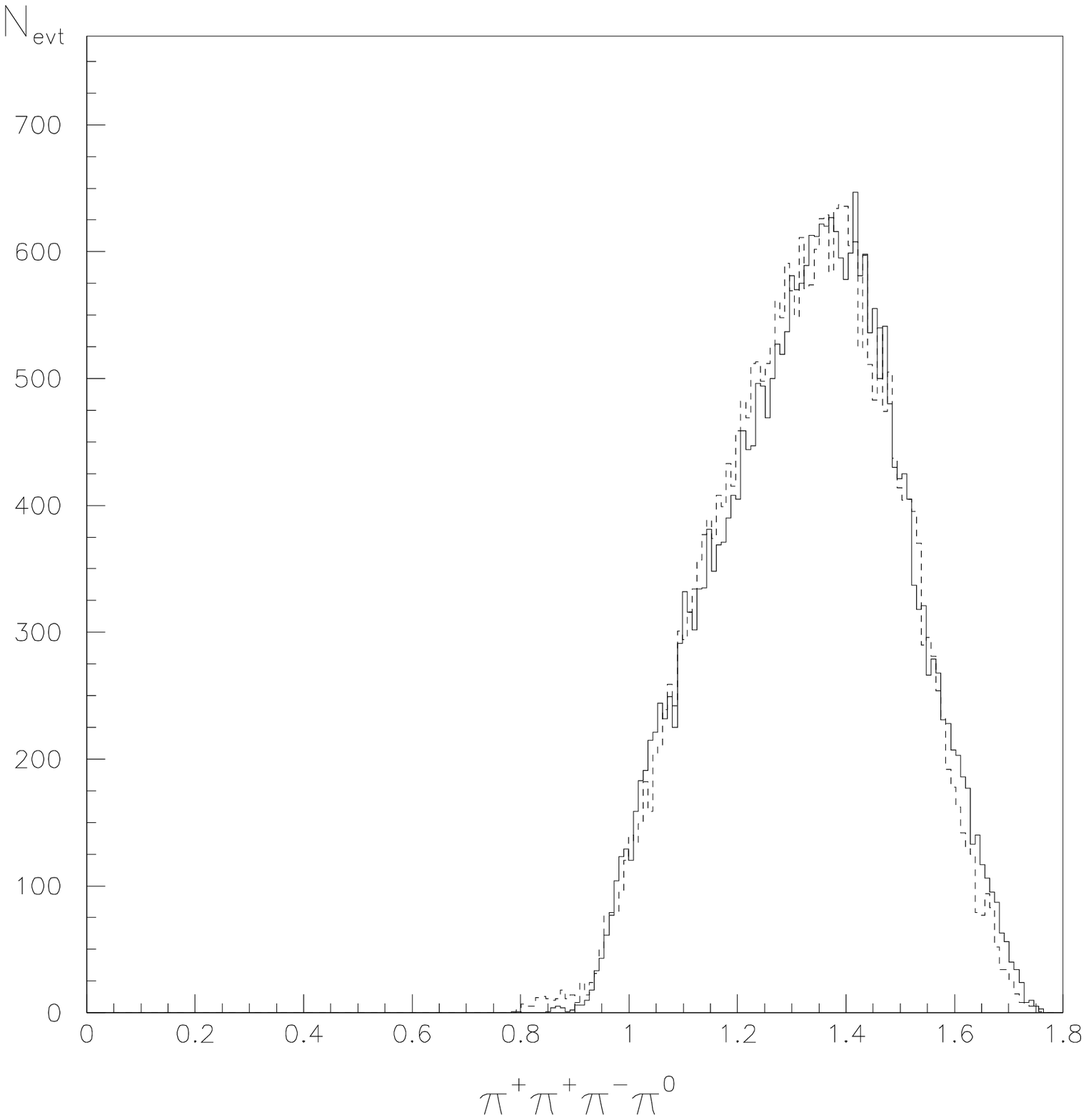,width=80mm,height=80mm}}}
\put(700, -1){\makebox(0,0)[lb]{\epsfig{file=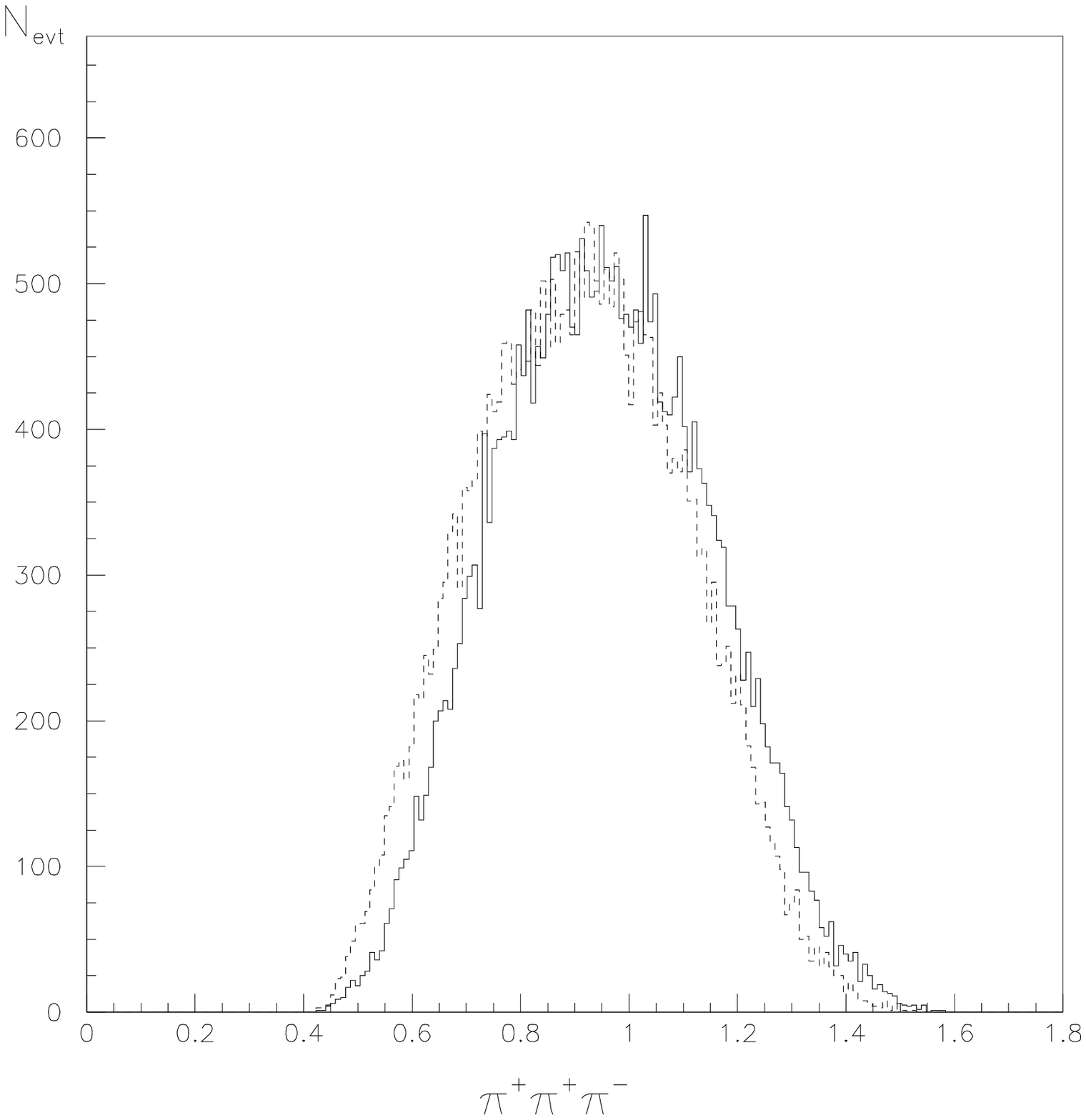,width=80mm,height=80mm}}}
\end{picture}
\caption
{\it The $\bar{\nu}_{\tau}\pi^{+}\pi^{+}\pi^{-}\pi^{0}$ channel.  
The  left-hand side plots the $\pi^{+}\pi^{+}\pi^{-}\pi^{0}$ invariant
mass distribution and the right-hand side is the $\pi^{+}\pi^{+}\pi^{-}$ 
invariant mass distribution. The continuous and dotted lines correspond to
the old CLEO and new Novosibirsk current respectively.}
\label{2pipi0pi1}
\end{figure}
\begin{figure}[!ht]
\setlength{\unitlength}{0.1mm}
\begin{picture}(1600,800)
\put( 375,750){\makebox(0,0)[b]{\large }}
\put(1225,750){\makebox(0,0)[b]{\large }}
\put(-20, -1){\makebox(0,0)[lb]{\epsfig{file=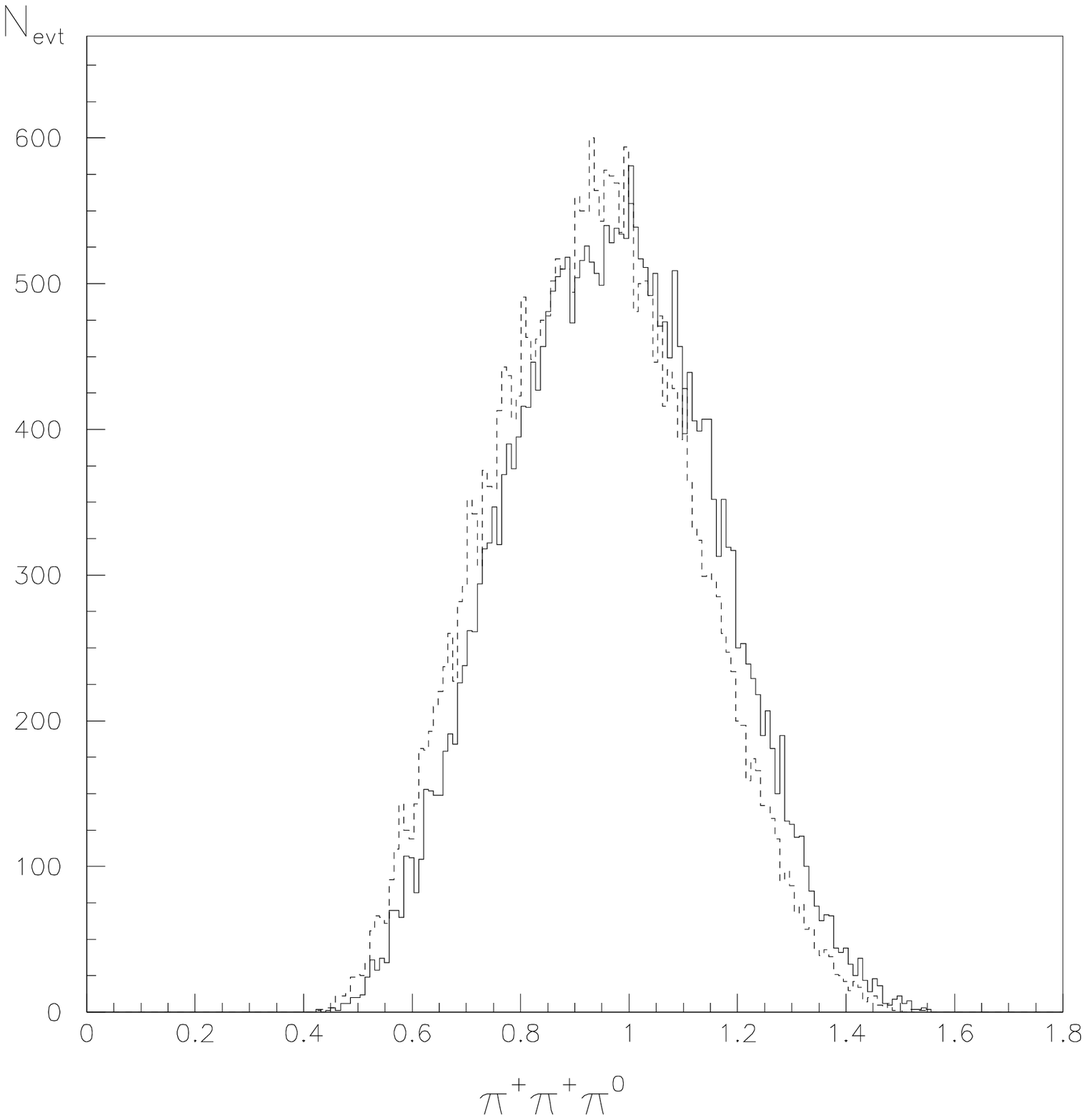,width=80mm,height=80mm}}}
\put(700, -1){\makebox(0,0)[lb]{\epsfig{file=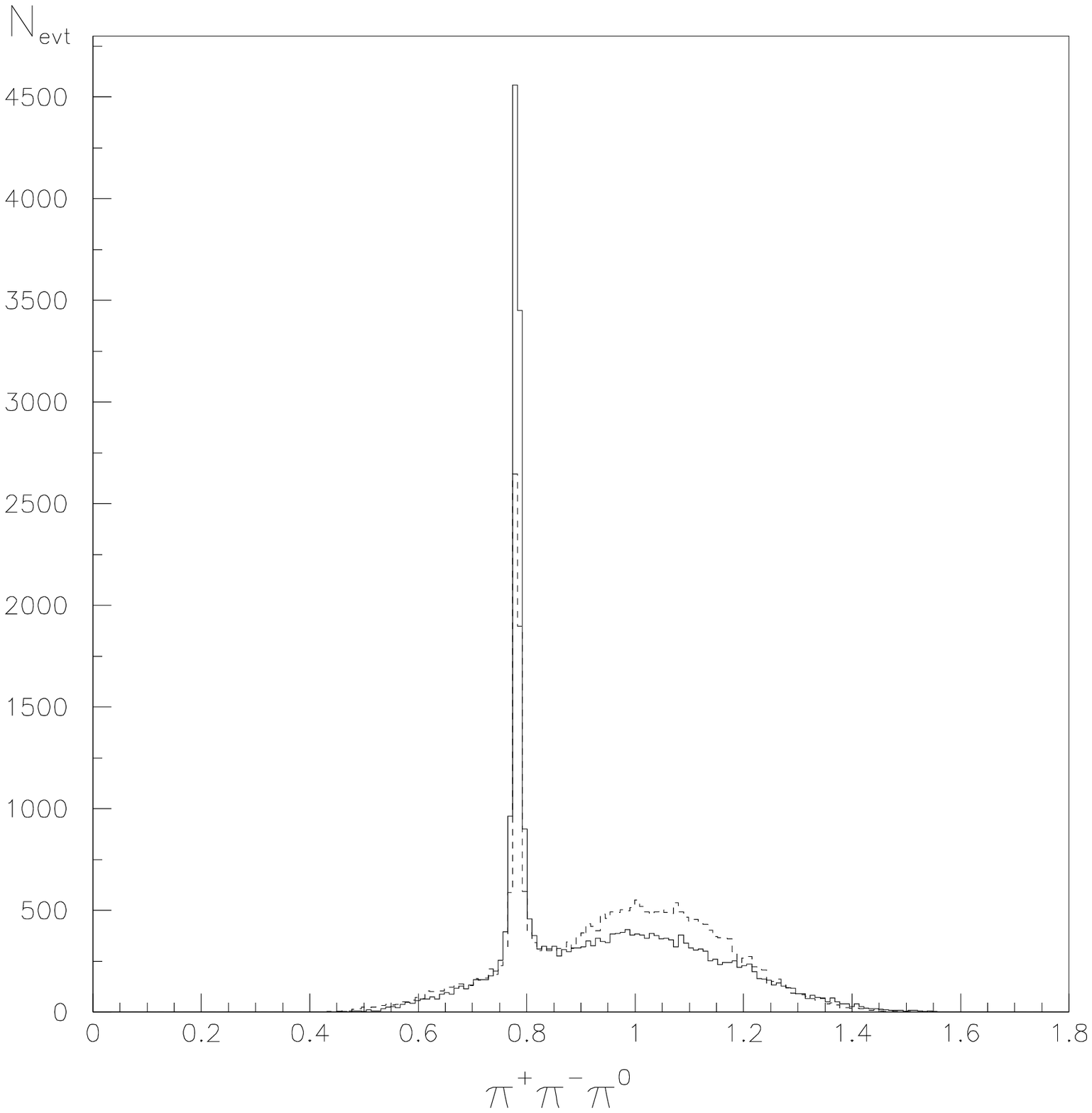,width=80mm,height=80mm}}}
\end{picture}
\caption 
{\it The $\bar{\nu}_{\tau}\pi^{+}\pi^{+}\pi^{-}\pi^{0}$ channel. The  
left-hand side plots the $\pi^{+}\pi^{+}\pi^{0}$ invariant
mass distribution and the right-hand side is the $\pi^{+}\pi^{-}\pi^{0}$ 
invariant mass distribution. The continuous and dotted lines correspond to
the old CLEO and new Novosibirsk current respectively.}
\label{2pipi0pi2}
\end{figure}
\begin{figure}[!ht]
\setlength{\unitlength}{0.1mm}
\begin{picture}(1600,800)
\put( 375,750){\makebox(0,0)[b]{\large }}
\put(1225,750){\makebox(0,0)[b]{\large }}
\put(-20, -1){\makebox(0,0)[lb]{\epsfig{file=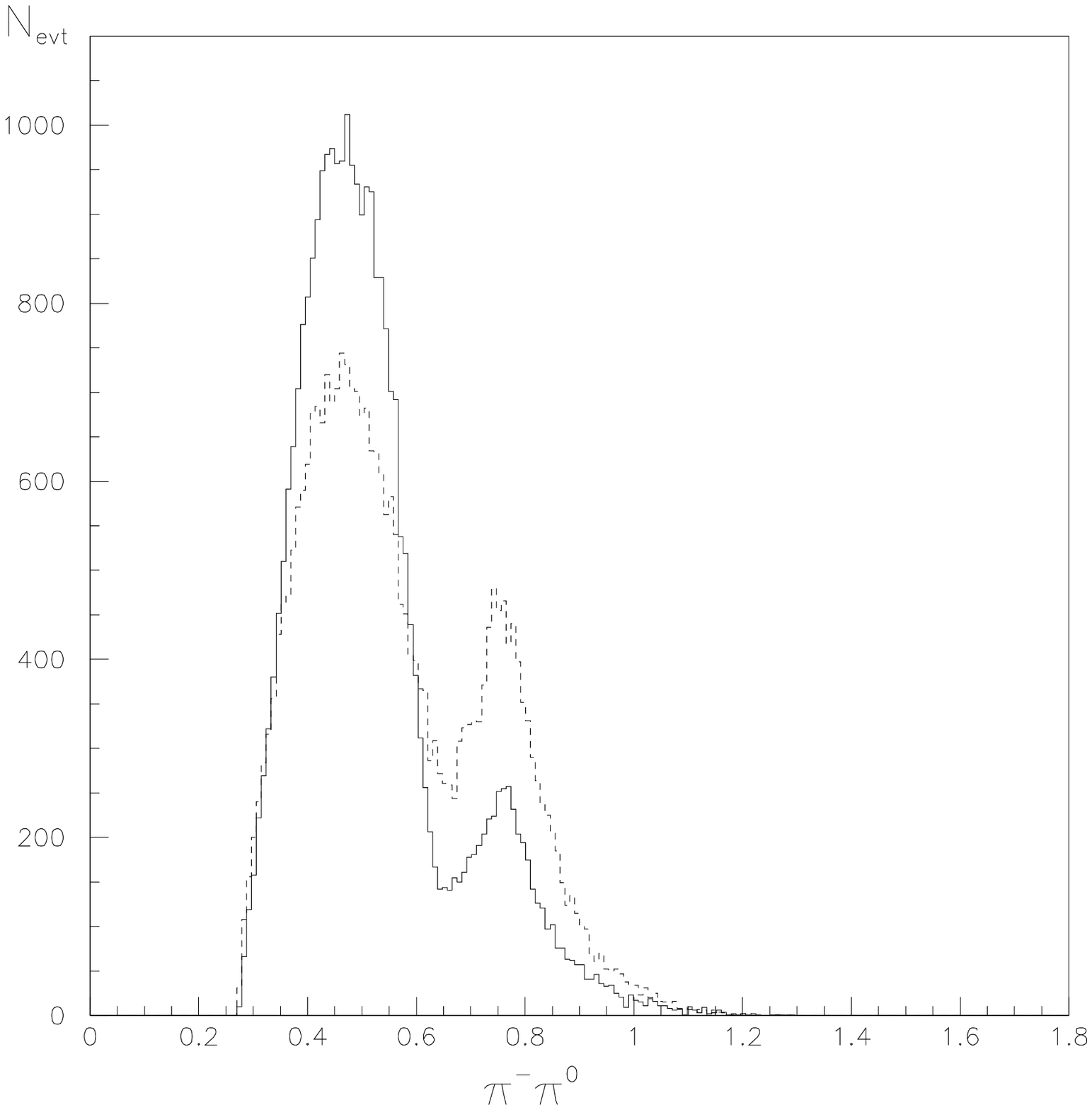,width=80mm,height=80mm}}}
\put(700, -1){\makebox(0,0)[lb]{\epsfig{file=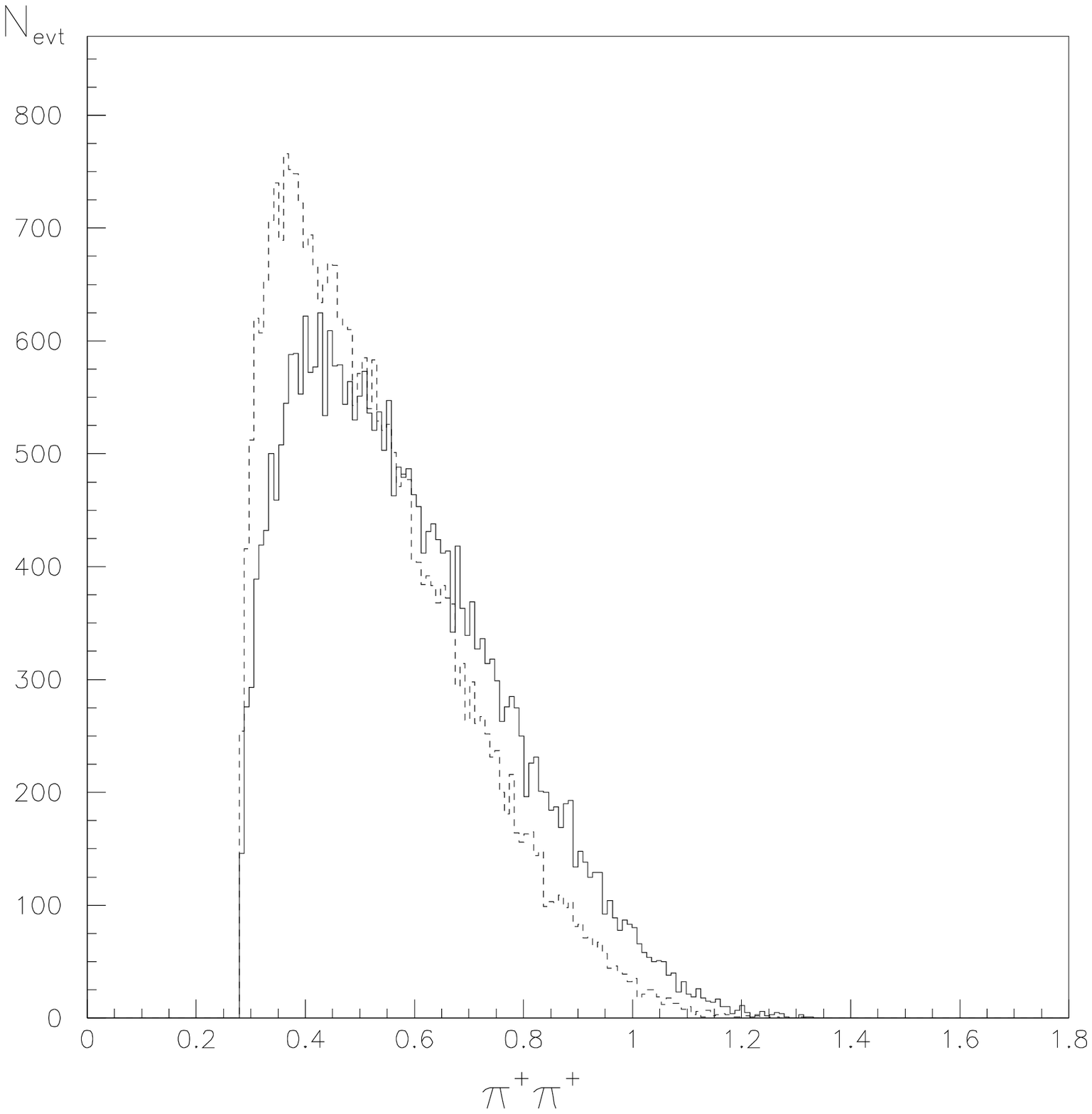,width=80mm,height=80mm}}}
\end{picture}
\caption 
{\it  The $\bar{\nu}_{\tau}\pi^{+}\pi^{+}\pi^{-}\pi^{0}$ channel. The  
left-hand side plots the $\pi^{-}\pi^{0}$ invariant
mass distribution and the right-hand side is the $\pi^{+}\pi^{+}$ invariant
mass distribution. The continuous and dotted lines correspond to the old CLEO 
and new Novosibirsk current respectivevly.}
\label{2pipi0pi3}
\end{figure}
\begin{figure}[!ht]
\setlength{\unitlength}{0.1mm}
\begin{picture}(1600,800)
\put( 375,750){\makebox(0,0)[b]{\large }}
\put(1225,750){\makebox(0,0)[b]{\large }}
\put(-20, -1){\makebox(0,0)[lb]{\epsfig{file=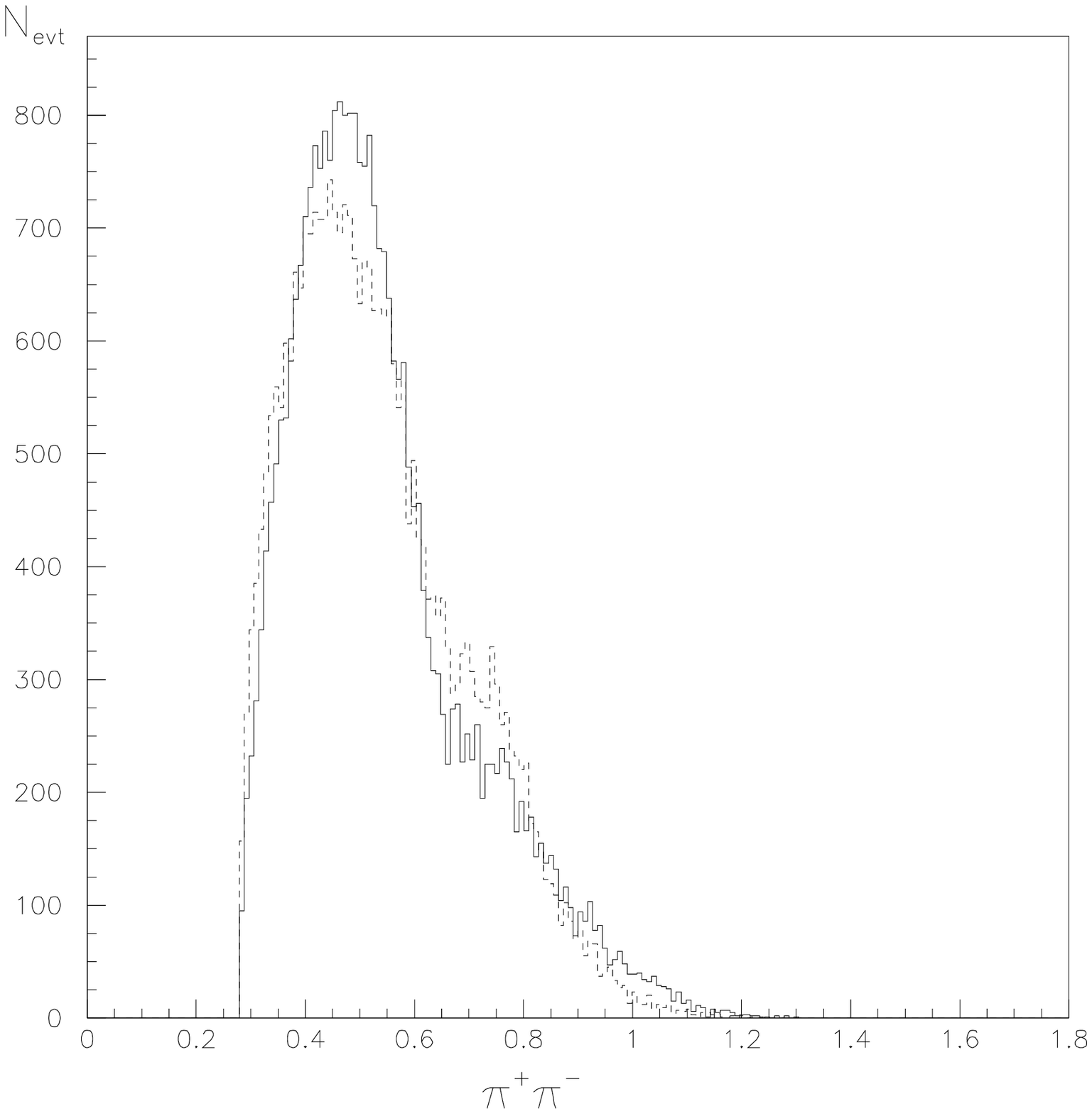,width=80mm,height=80mm}}}
\put(700, -1){\makebox(0,0)[lb]{\epsfig{file=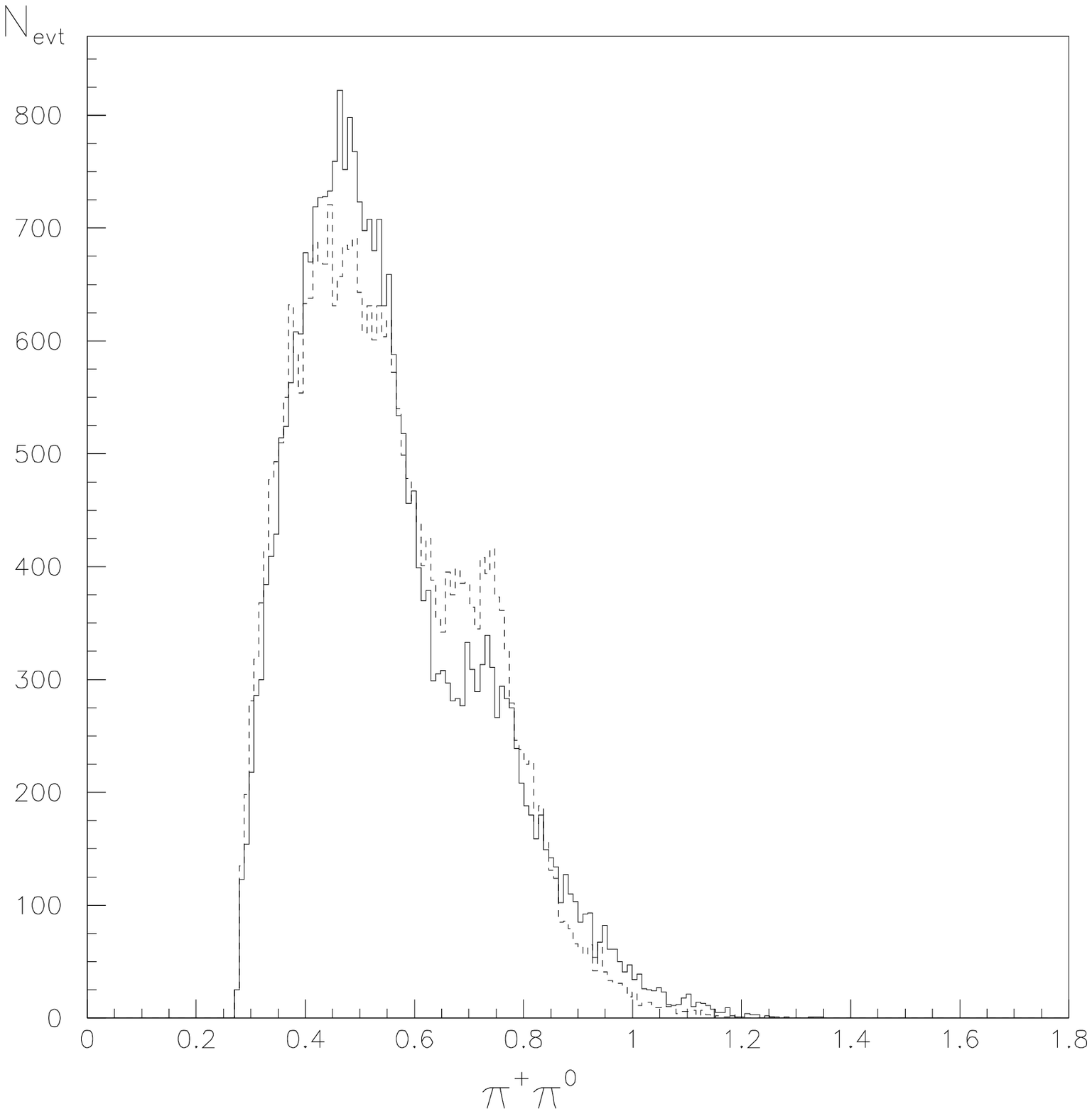,width=80mm,height=80mm}}}
\end{picture}
\caption 
{\it  The $\bar{\nu}_{\tau}\pi^{+}\pi^{+}\pi^{-}\pi^{0}$ channel. The 
left-hand side plots the $\pi^{+}\pi^{-}$ invariant
mass distribution and the right-hand side is the $\pi^{+}\pi^{0}$ invariant
mass distribution.
The continuous and dotted lines correspond to the old CLEO and new Novosibirsk
current respectively.}

\label{2pipi0pi4}
\end{figure}

Once we became  confident in the correct implementation of the new 
Novosibirsk model, we have decided to  compare
its predictions to another one. For that purpose we have chosen the model 
tuned to the 1998 CLEO data 
because it is still quite often used.

In this section we collect predictions for all possible invariant mass 
distributions for the decay channels: 
$\tau\rightarrow\bar{\nu}_{\tau}\pi^{+}\pi^{-}\pi^{+}\pi^{0}$ and 
$\tau\rightarrow\bar{\nu}_{\tau}\pi^{+}\pi^{0}\pi^{0}\pi^{0}$ .
In Figs.~\ref{3pi0pi1} to~\ref{2pipi0pi4} 
we compare predictions of our new Novosibirsk current with the older 
CLEO current \cite{Pions1,test_cleo1,test_cleo2,test_cleo3 },
both implemented in the {\tt TAUOLA} Monte Carlo library as options: \\

$\bullet$ $\tau^{+}\rightarrow\bar{\nu}_{\tau}\pi^{+}\pi^{0}\pi^{0}\pi^{0}$
channel \\

 In the first part of this chapter the invariant mass distributions for 
$\tau^{+}\rightarrow\bar{\nu}_{\tau}\pi^{+}\pi^{0}\pi^{0}\pi^{0}$
channel are shown. In Fig.~\ref{3pi0pi1} we show the invariant
 mass distribution for $\pi^{0}\pi^{0}\pi^{0}\pi^{+}$ (left-hand side plot)
 and $\pi^{0}\pi^{0}\pi^{0}$ (right-hand side plot) systems. 
In Fig.~\ref{3pi0pi2} we show the invariant
 mass distribution for $\pi^{0}\pi^{0}\pi^{+}$ (left-hand side plot)
 and $\pi^{0}\pi^{0}$ (right-hand side plot) systems, 
and in Fig.~\ref{3pi0pi3} the invariant
 mass distribution for $\pi^{0}\pi^{+}$ system. 
In all plots the continuous and dotted lines correspond to the old 
(1998) CLEO and new Novosibirsk current.\\

$\bullet$ $\tau^{+}\rightarrow\bar{\nu}_{\tau}\pi^{+}\pi^{-}\pi^{+}\pi^{0}$
channel \\

In the second part of this chapter 
the invariant mass distributions for 
$\tau^{+}\rightarrow\bar{\nu}_{\tau}\pi^{+}\pi^{-}\pi^{+}\pi^{0}$
channel are shown. 
In Fig.~\ref{2pipi0pi1} we show the invariant
 mass distribution for $\pi^{+}\pi^{+}\pi^{-}\pi^{0}$  (left-hand side plot)
 and $\pi^{+}\pi^{+}\pi^{-}$ (right-hand side plot) systems. 
In Fig.~\ref{2pipi0pi2} we show the invariant
 mass distribution for $\pi^{+}\pi^{+}\pi^{0}$ (left-hand side plot)
 and $\pi^{+}\pi^{-}\pi^{0}$ (right-hand side plot) systems. 
In Fig.~\ref{2pipi0pi3} we show the invariant
 mass distribution for $\pi^{+}\pi^{0}$ (left-hand side plot)
 and  $\pi^{+}\pi^{+}$ (right-hand side plot) systems.
In Fig.~\ref{2pipi0pi4} we show the invariant
 mass distribution for $\pi^{+}\pi^{-}$ (left-hand side plot)
 and  $\pi^{+}\pi^{0}$ (right-hand side plot) systems. 

In some cases agreement is only qualitatively correct, but one should 
have in mind rather limited data samples available at that 
time\footnote{ The largest differences
are present   in  the
$\tau^{+}\rightarrow\bar{\nu}_{\tau}\pi^{+}\pi^{-}\pi^{+}\pi^{0}$ decay
channel. They  can be substantially diminished, if the $\omega\pi$ 
contribution to the current used in the CLEO model is appropriately reduced 
\cite{Tomek} to match the present measurements.}.
We  expect the  Novosibirsk model to represent a substantial
improvement over the old one.
 
\section{ Summary}
\vskip 0.3 cm
The new parameterization of  $4\pi$ form factors for the
 {\tt TAUOLA} package is now
available. The form factors are completely defined from the information 
in the paper.  The particular strength of the  model used in their definition
relies on its success in describing high statistics low energy $e^+e^-$ data
at $\sqrt{s} <$ 1.4 GeV (future experiments at the upgraded collider
VEPP-2000 will extend the energy range to the $\tau$ lepton mass \cite{v2000}). 
The CVC hypothesis has been instrumental in constructing predictions
for the $\tau$ decays. The correctness of the code
was checked by its comparison to the other generator based  on the same data. 
Results obtained from our  new form factors can be now
used for comparisons with other models
and $\tau$ lepton decay data directly  
under conditions of any present or future experiment.\\ \\

\vskip 0.3 cm
\noindent {\large \bf Acknowledgements}\\

Two of the authors (T.P. and M.W.) are very grateful for the warm 
hospitality extended to them by 
the Budker Institute of Nuclear Physics in Novosibirsk,
where the sizable part of this work was performed. \\

\newpage
\vskip 0.5 cm
 \begin{center}
{ \Large \bf Appendix A }
 
\end{center}
\vskip 0.5 cm
\begin{center}
{\large \it Feynman diagrams for $\tau^{+}$ decays into $4\pi$}
\end{center}
\vskip 0.3 cm

Taking into account the quantum numbers of all intermediate 
and final states, 
see Table~\ref{numbers}, the following  Feynman
diagrams can be written for the $\tau^{+}$
decay into $\bar{\nu}_{\tau}\pi^{+} \pi^{0}\pi^{0}\pi^{0}$ and  
$\bar{\nu}_{\tau}\pi^{+}\pi^{-}\pi^{+}\pi^{0}$, if  $a_{1}(1260)\pi$ and  
$\omega\pi$ intermediate states are assumed.

\begin{table}[!h]
\newcommand{\lstrut}{{$\strut\atop\strut$}}
  \caption {\em The quantum numbers of mesons. 
\label{numbers}}
\vspace{2mm}
\begin{center}
\begin{tabular}{|c|c|c|c|} \hline \hline 
 Intermediate &        &  \\               
  state                         & $I^{G}$  & $ ~~J^{PC}$   \\ \hline \hline
 $\rho(770)$ ($\tilde{\rho}$)    & $1^{+}$  & $ ~~1^{--}$   \\ \hline
 $\pi^{\pm}$                & $1^{-}$  & $ 0^{- }$   \\ \hline
 $\pi^{0}$                  & $1^{-}$  & $ ~~0^{-+}$   \\ \hline
 $a_{1}(1260)$              & $1^{-}$  & $ ~~1^{++}$   \\ \hline
 $\omega(782)$                   & $0^{-}$  & $ ~~1^{--}$   \\ \hline
 $\sigma$                   & $0^{+}$  & $ ~~0^{++}$    \\ \hline \hline
\end{tabular}
\end{center}
\end{table}
\begin{figure}[!ht]
\setlength{\unitlength}{0.1mm}
\begin{picture}(1600,800)
\put( 375,750){\makebox(0,0)[b]{\large }}
\put(1225,750){\makebox(0,0)[b]{\large }}
\put(80, -600){\makebox(0,0)[lb]{\epsfig{file=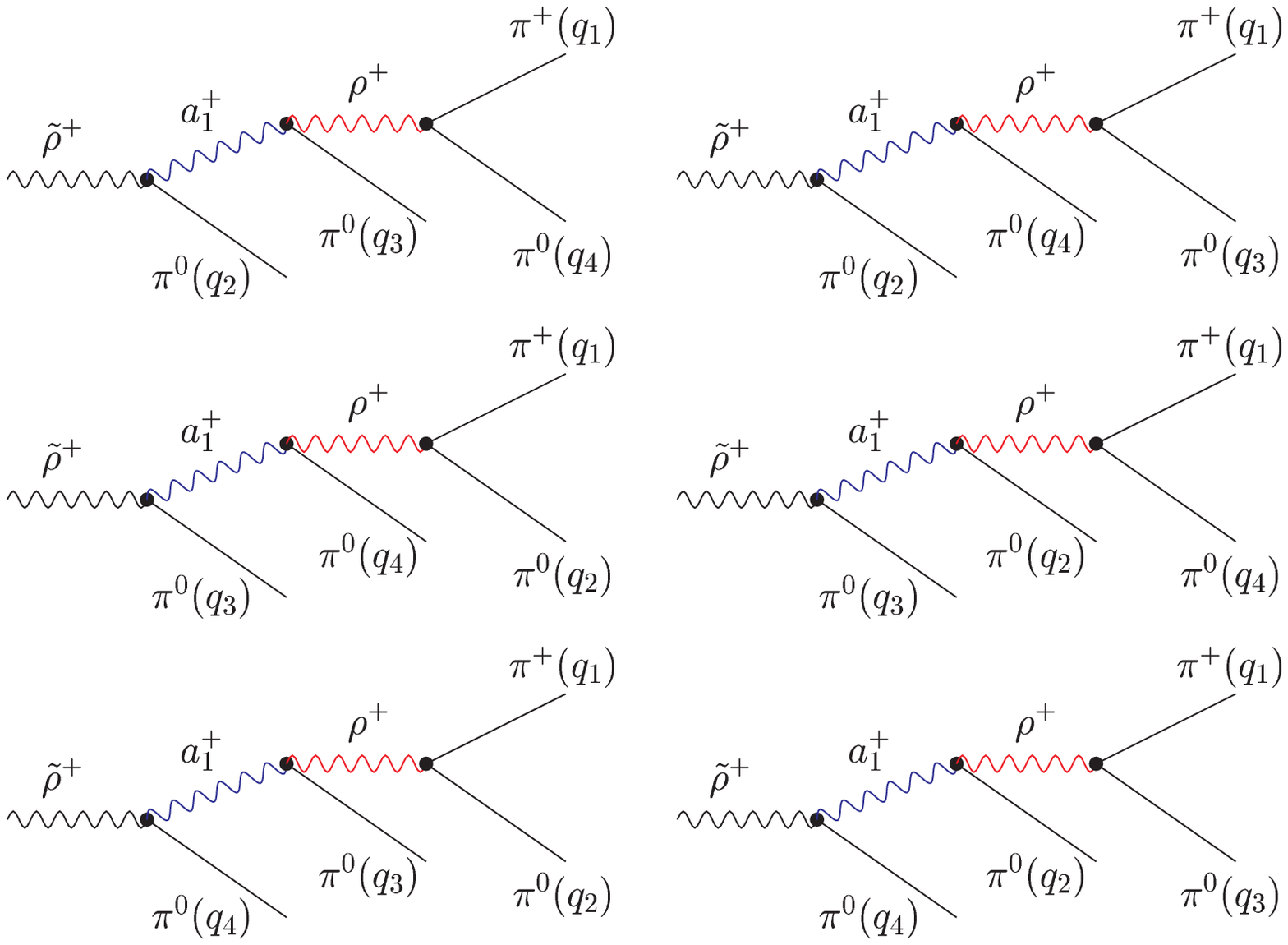,width=130mm,height=130mm}}}
\end{picture}
\caption 
{\it Diagrams for the $\tau^{+}\rightarrow\bar{\nu}_{\tau}
\pi^{+}(q_{1})\pi^{0}(q_{2})\pi^{0}(q_{3})\pi^{0}(q_{4})$ decay
via the $\tilde{\rho}^{+}\rightarrow a_{1}\pi\rightarrow\rho\pi\pi
\rightarrow\pi\pi\pi\pi$
 intermediate states. }
\label{diagram1}
\end{figure}

\begin{figure}[!ht]
\setlength{\unitlength}{0.1mm}
\begin{picture}(1600,800)
\put( 375,750){\makebox(0,0)[b]{\large }}
\put(1225,750){\makebox(0,0)[b]{\large }}
\put(80, -600){\makebox(0,0)[lb]{\epsfig{file=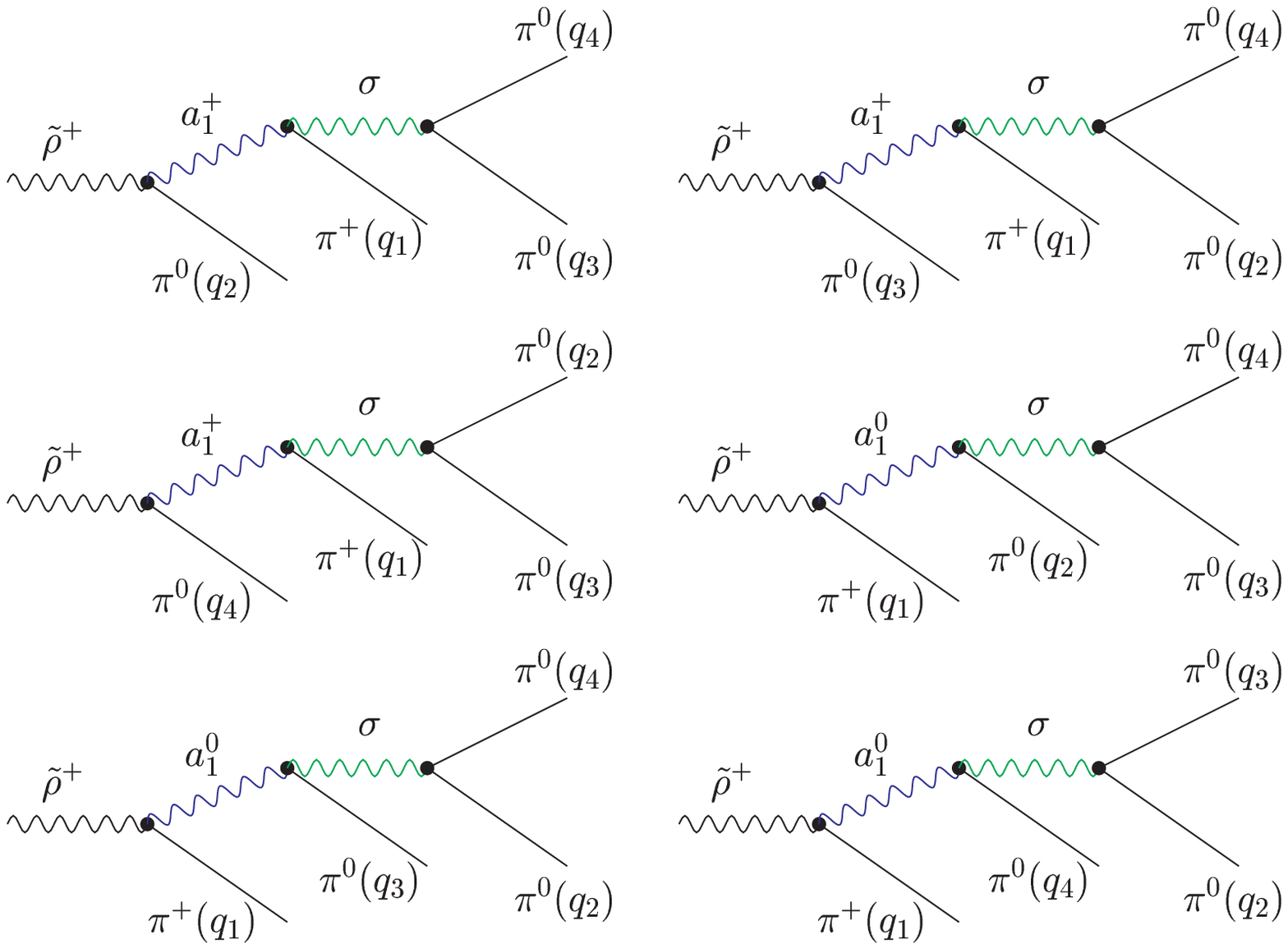,width=130mm,height=130mm}}}
\end{picture}
\caption 
{\it Diagrams for the $\tau^{+}\rightarrow\bar{\nu}_{\tau}
\pi^{+}(q_{1})\pi^{0}(q_{2})\pi^{0}(q_{3})\pi^{0}(q_{4})$ decay channel
via the $\tilde{\rho}^{+}\rightarrow a_{1}\pi\rightarrow
\sigma\pi\pi\rightarrow \pi\pi\pi\pi$ intermediate states.}
\label{diagram2}
\end{figure}

\begin{figure}[!ht]
\setlength{\unitlength}{0.1mm}
\begin{picture}(1600,800)
\put( 375,750){\makebox(0,0)[b]{\large }}
\put(1225,750){\makebox(0,0)[b]{\large }}
\put(80, -600){\makebox(0,0)[lb]{\epsfig{file=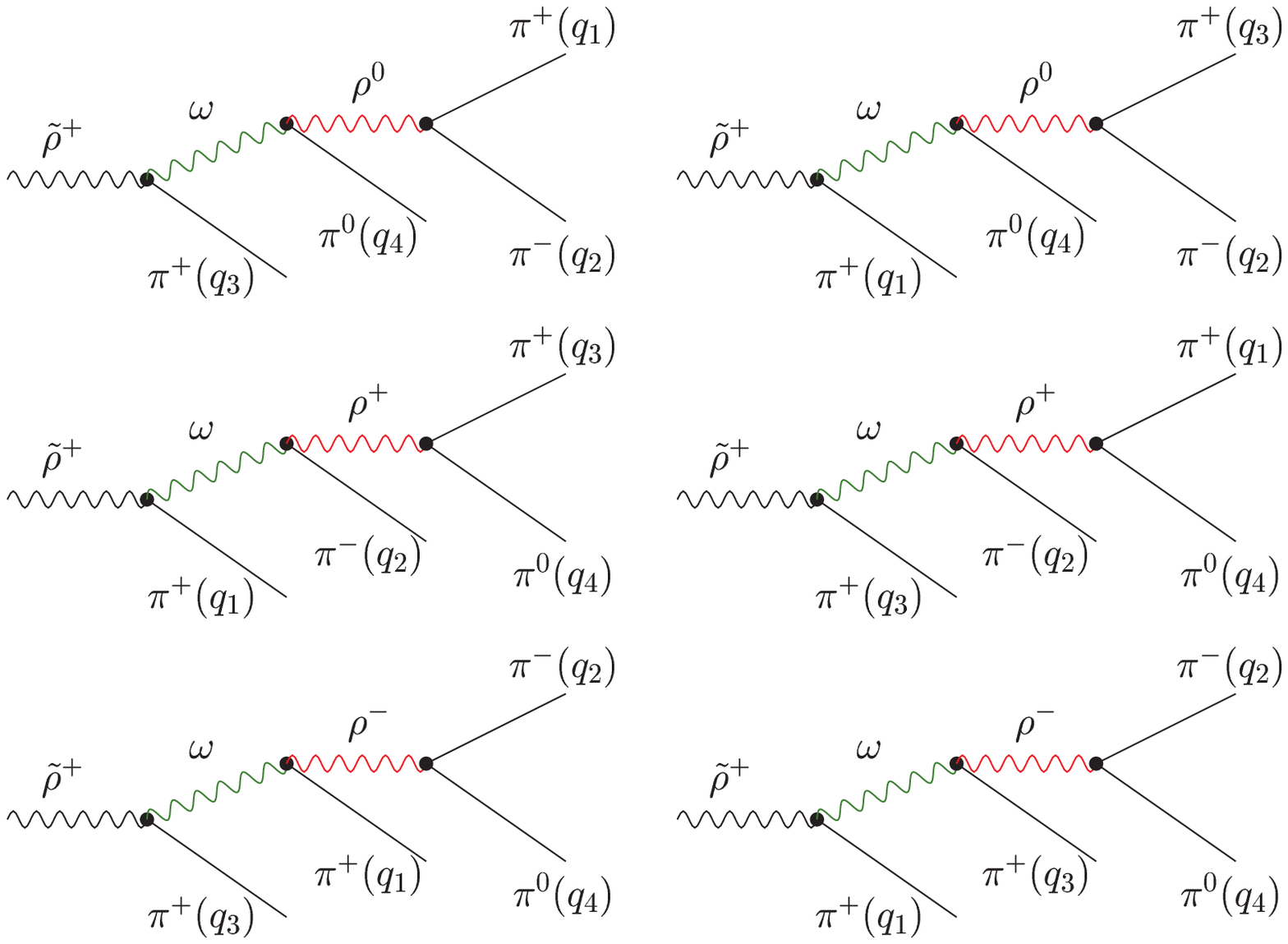,width=130mm,height=130mm}}}
\end{picture}
\caption 
{\it Diagrams for the $\tau^{+}\rightarrow\bar{\nu}_{\tau}
\pi^{+}(q_{1})\pi^{-}(q_{2})\pi^{+}(q_{3})\pi^{0}(q_{4})$ decay
via the $\tilde{\rho}^{+}\rightarrow\omega\pi\rightarrow\rho\pi\pi\rightarrow
\pi\pi\pi\pi$ intermediate states.}
\label{diagram3}
\end{figure}
\begin{figure}[!ht]
\setlength{\unitlength}{0.1mm}
\begin{picture}(1600,800)
\put( 375,750){\makebox(0,0)[b]{\large }}
\put(1225,750){\makebox(0,0)[b]{\large }}
\put(80, -600){\makebox(0,0)[lb]{\epsfig{file=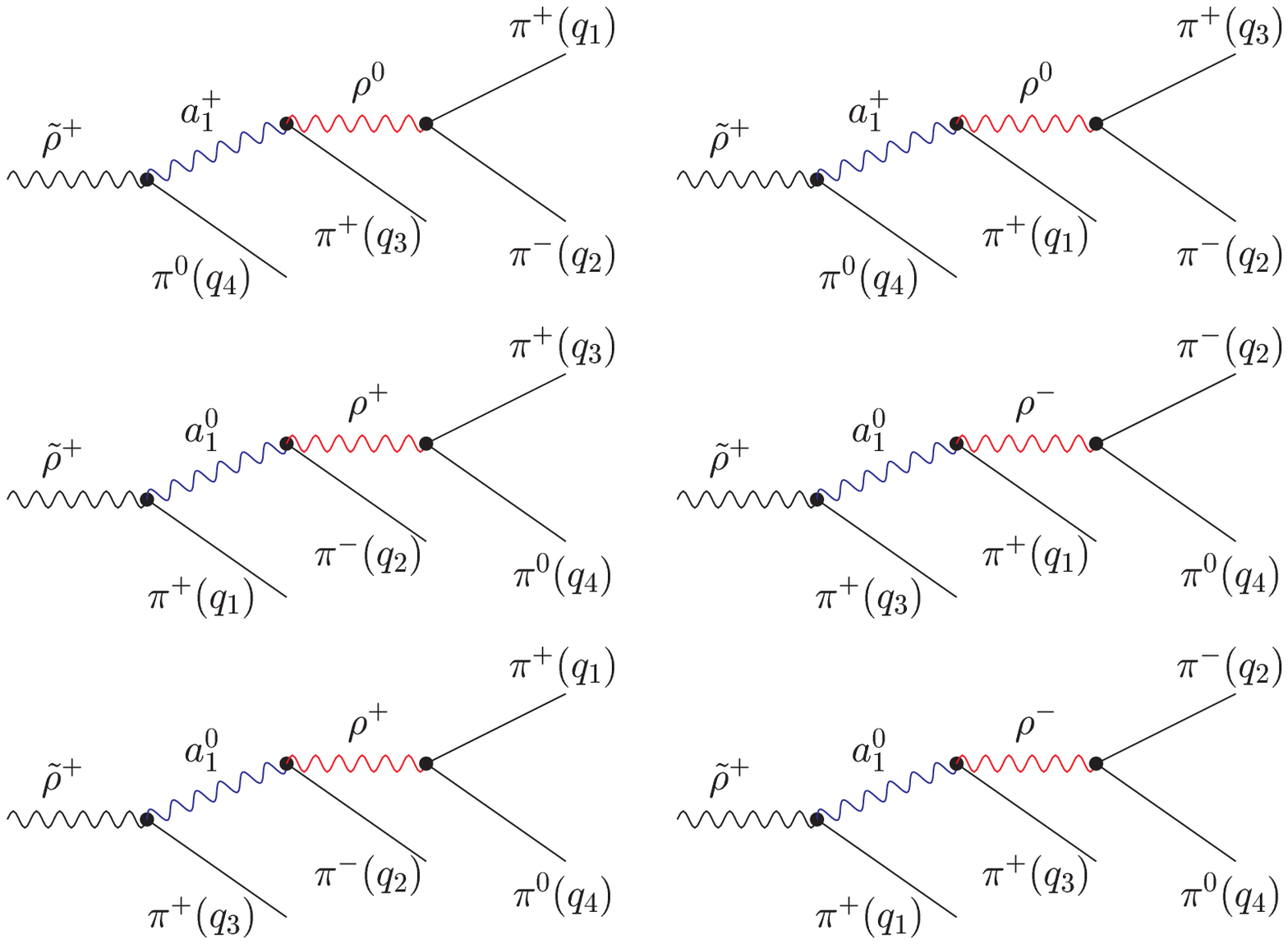,width=130mm,height=130mm}}}
\end{picture}
\caption 
{\it Diagrams for the $\tau^{+}\rightarrow\bar{\nu}_{\tau}
\pi^{+}(q_{1})\pi^{-}(q_{2})\pi^{+}(q_{3})\pi^{0}(q_{4})$ decay
via the $\tilde{\rho}^{+}\rightarrow a_{1}\pi\rightarrow\rho\pi\pi
\rightarrow\pi\pi\pi\pi$
 intermediate states.}
\label{diagram4}
\end{figure}
\begin{figure}[!ht]
\setlength{\unitlength}{0.1mm}
\begin{picture}(1600,800)
\put( 375,750){\makebox(0,0)[b]{\large }}
\put(1225,750){\makebox(0,0)[b]{\large }}
\put(80, -600){\makebox(0,0)[lb]{\epsfig{file=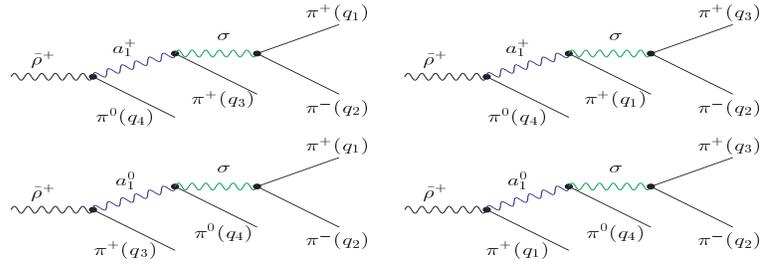,width=130mm,height=130mm}}}
\end{picture}
\caption 
{\it Diagrams for the $\tau^{+}\rightarrow\bar{\nu}_{\tau}
\pi^{+}(q_{1})\pi^{-}(q_{2})\pi^{+}(q_{3})\pi^{0}(q_{4})$ decay
via the $\tilde{\rho}^{+}\rightarrow a_{1}\pi\rightarrow
\sigma\pi\pi\rightarrow \pi\pi\pi\pi$ intermediate states }
\label{diagram5}
\end{figure}

\newpage
\hspace{1 mm}
\vspace{10 cm}

\begin{center}
{ \Large \bf Appendix B}
\end{center}
\vskip 0.5 cm
\begin{center}
{\large \it Tables of numerical values for functions $G(Q^2)$}
\end{center}
\vskip 0.3 cm

Tables of numerical values for the 
$G_{\pi^{+}\pi^{0}\pi^{0}\pi^{0}}(Q^{2})$, $G_{\pi^{+}\pi^{-}\pi^{+}\pi^{0}}
(Q^{2})$, and $G^\omega_{\pi^{+}\pi^{-}\pi^{+}\pi^{0}}(Q^{2})$ functions 
at 96 energy points are given.
In Table~\ref{fits}.A  the 
$G_{\pi^{+}\pi^{0}\pi^{0}\pi^{0}}(Q^{2})$ function for the 
$\pi^{+}\pi^{0}\pi^{0}\pi^{0}$ decay channel -  part of the current
dominated by the  $a_{1}\pi$ intermediate state is given. 
In Table~\ref{fits}.B the 
$G_{\pi^{+}\pi^{-}\pi^{+}\pi^{0}}(Q^{2})$ function for the 
$\pi^{+}\pi^{-}\pi^{+}\pi^{0}$  decay channel - part of the current
dominated by  the $a_{1}\pi$
intermediate state is given , and finally in Table~\ref{fits}.C the 
$G^{\omega}_{\pi^{+}\pi^{-}\pi^{+}\pi^{0}}(Q^{2})$  for the 
$\pi^{+}\pi^{-}\pi^{+}\pi^{0}$ decay channel -
 and  $\omega\pi$ intermediate state.
\begin{table}[!h]
\begin{center}
\setlength{\unitlength}{1cm}
\begin{picture}(20,15)
\put( 0,7){\makebox(0,0)[lb]{\epsfig{file=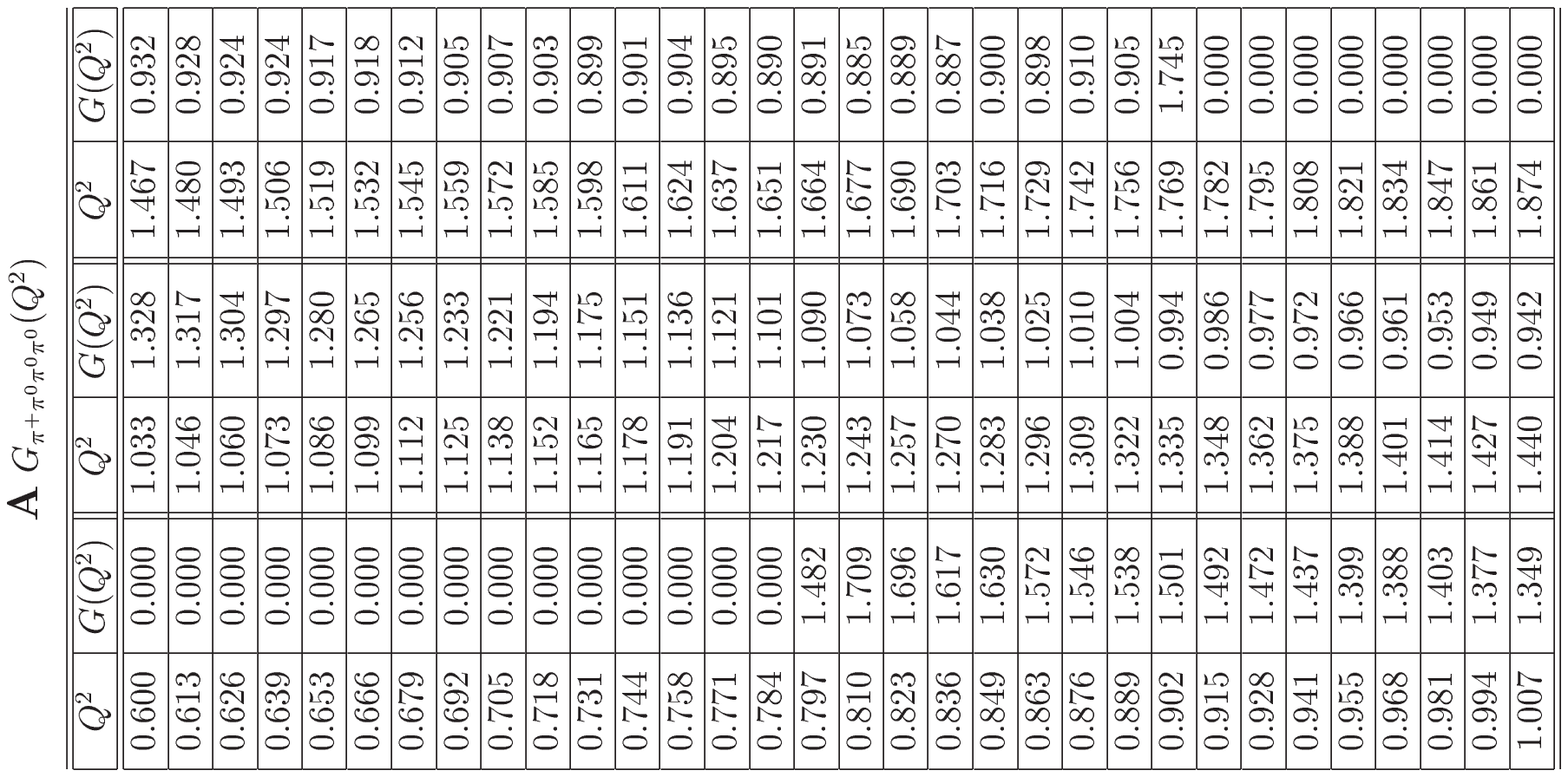,width=20cm,height=11cm,angle=90}}}
\put( 0, 2){\makebox(0,0)[lb]{\epsfig{file=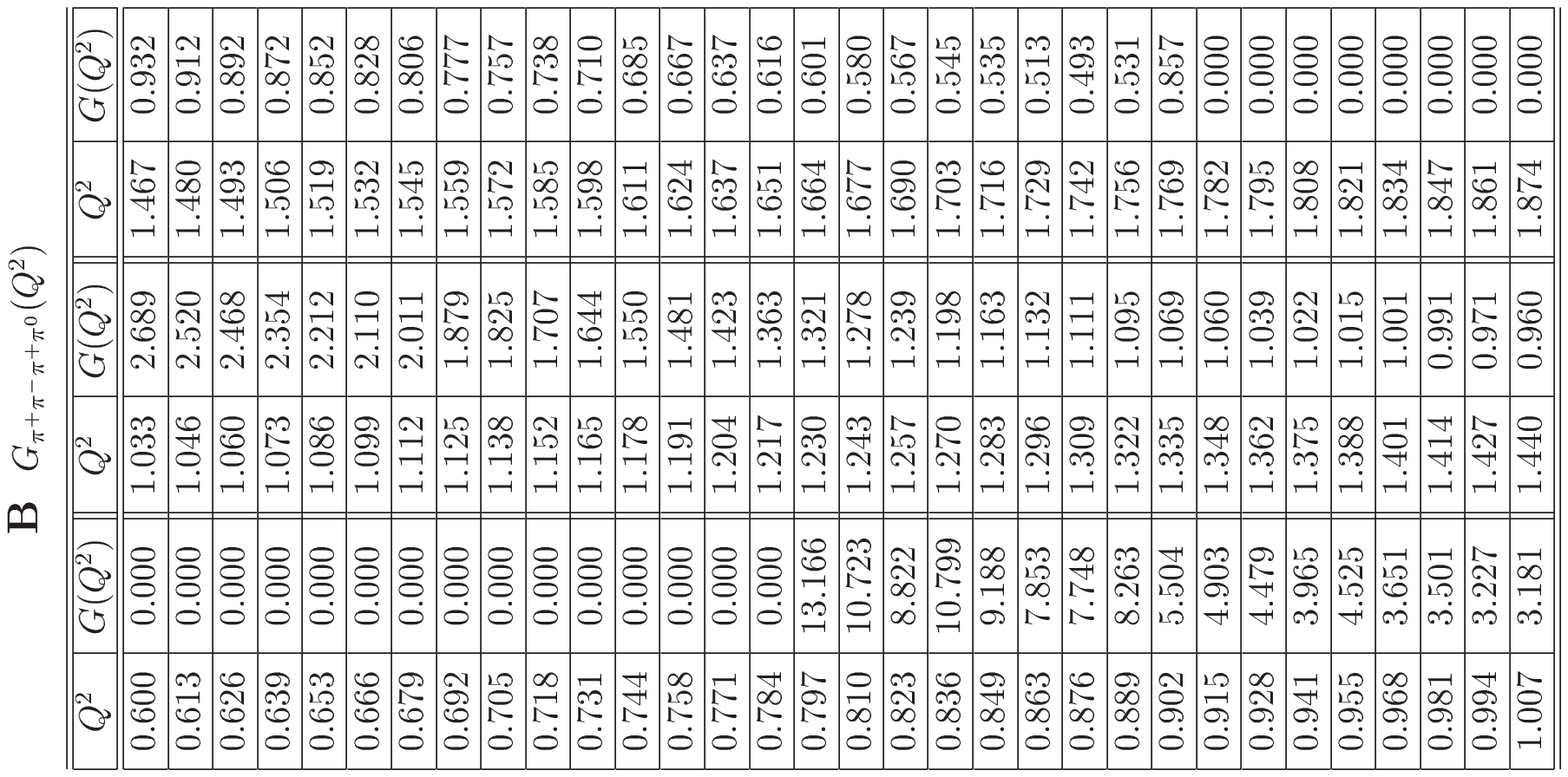,width=20cm,height=11cm,angle=90}}}
\put( 0, -3){\makebox(0,0)[lb]{\epsfig{file=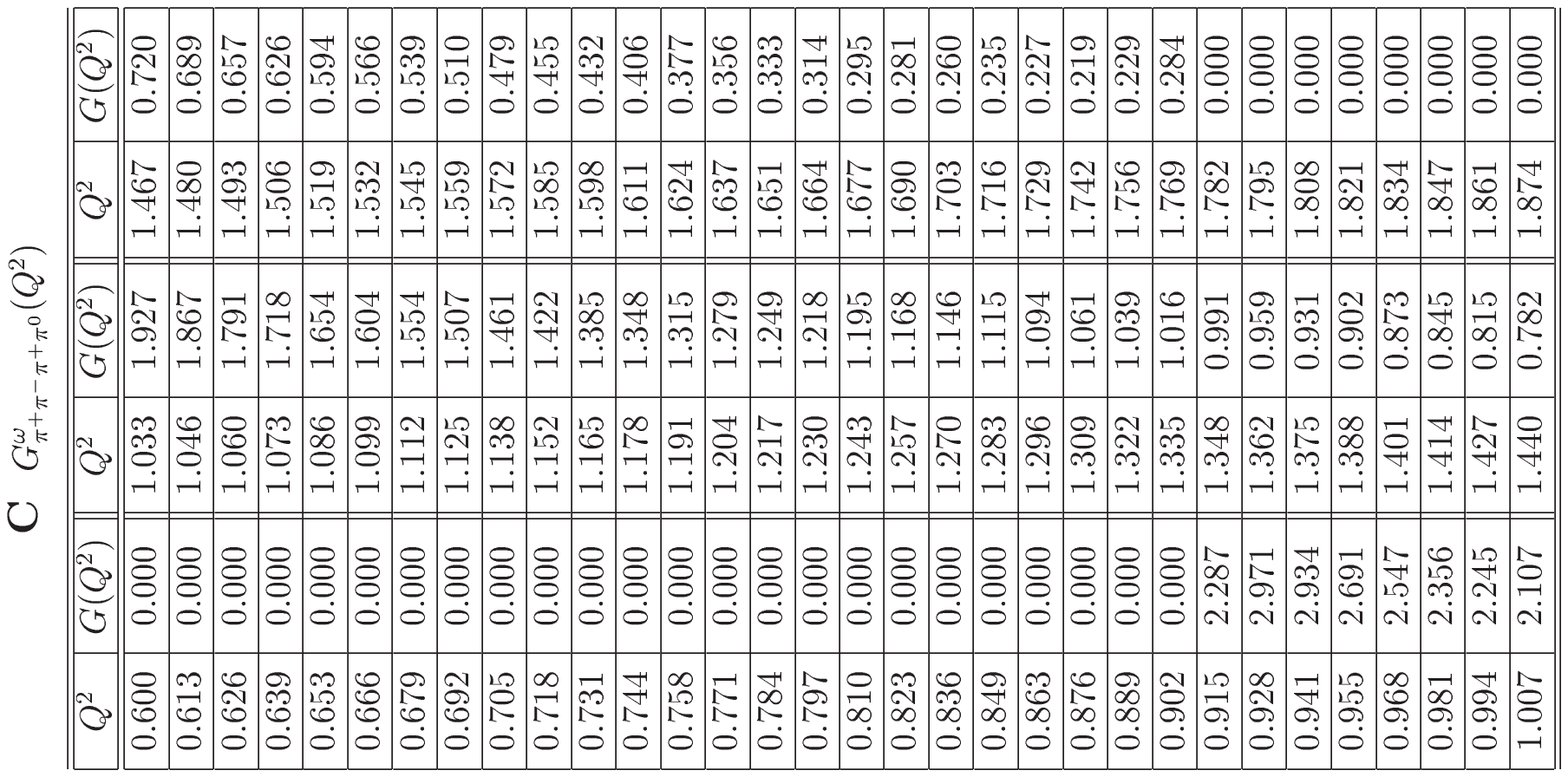,width=20cm,height=11cm,angle=90}}}
\end{picture}
\caption{\em The tables of numerical values for the following functions: 
\newline {\bf (A)}
$G_{\pi^{+}\pi^{0}\pi^{0}\pi^{0}}(Q^{2})$ used in the 
$\pi^{+}\pi^{0}\pi^{0}\pi^{0}$ decay channel 
and $a_{1}\pi$ intermediate state.
\newline {\bf (B)}
$G_{\pi^{+}\pi^{-}\pi^{+}\pi^{0}}(Q^{2})$ used in the 
$\pi^{+}\pi^{-}\pi^{+}\pi^{0}$ decay channel 
and $a_{1}\pi$
intermediate state.
\newline {\bf (C)} 
$G^{\omega}_{\pi^{+}\pi^{-}\pi^{+}\pi^{0}}(Q^{2})$ used in the
$\pi^{+}\pi^{-}\pi^{+}\pi^{0}$ decay channel 
and $\omega\pi$ intermediate state.
}
\label{fits}
\end{center}
\end{table}

\end{document}